\documentclass[fleqn,usenatbib]{mnras}

\usepackage{newtxtext,newtxmath}
\usepackage[T1]{fontenc}

\DeclareRobustCommand{\VAN}[3]{#2}
\let\VANthebibliography\thebibliography
\def\thebibliography{\DeclareRobustCommand{\VAN}[3]{##3}\VANthebibliography}
\usepackage{bm}

\usepackage{graphicx}	% Including figure files
\usepackage{amsmath}	% Advanced maths commands
\usepackage{algorithm}
\usepackage{algpseudocode}
\usepackage{caption}
\usepackage{subcaption}
\usepackage{multicol}
\usepackage{paralist}
\usepackage{natbib}
\usepackage{xcolor}
\definecolor{pink}{rgb}{0.96, 0.76, 0.76}
\definecolor{aqua}{rgb}{0.22, 0.96, 0.93}
\definecolor{darkred}{rgb}{0.76, 0.23, 0.13}

\title[Normalizing flows to denoise the Milky Way]{Denoising Milky Way stellar survey data with normalizing flow models}

\author[Z. Yan and J. L. Sanders]{
Ziyang Yan\thanks{E-mail: ziyang.yan.17@ucl.ac.uk (ZY)} and 
Jason~L.~Sanders
\\
Department of Physics and Astronomy, University College London, London WC1E 6BT, UK
}

\date{Accepted XXX. Received YYY; in original form ZZZ}

\pubyear{\the\year{}}

\begin{document}
\label{firstpage}
\pagerange{\pageref{firstpage}--\pageref{lastpage}}
\maketitle

\begin{abstract}
The \emph{Gaia} dataset has revealed many intricate Milky Way substructures in exquisite detail, including moving groups and the phase spiral.
Precise characterisation of these features and detailed comparisons to theoretical models require engaging with \emph{Gaia}'s heteroscedastic noise model, particularly in more distant parts of the Galactic disc and halo.
We propose a general, novel machine-learning approach using normalizing flows for denoising density estimation, with particular focus on density estimation from stellar survey data such as that from \emph{Gaia}.
Normalizing flows transform a simple base distribution into a complex target distribution through bijective transformations resulting in a highly expressive and flexible model. The denoising is performed using importance sampling.
We demonstrate that this general procedure works excellently on \emph{Gaia} data by reconstructing detailed local velocity distributions artificially corrupted with noise.
For example, we show the multiple branches of the Hercules stream and the phase-space spiral can both be well captured by our model.
We discuss hyperparameter choice to optimally recover substructure and compare our approach to extreme deconvolution.
The model therefore promises to be a robust tool for studying the Milky Way's kinematics in Galactic locations where the noise from \emph{Gaia} is significant.
\end{abstract}


%%%%%%%%%%%%%%%%%%%%%%%%%%%%%%%%%%%%%%%%%%%%%%%%%%

%%%%%%%%%%%%%%%%% BODY OF PAPER %%%%%%%%%%%%%%%%%%

\section{Introduction}\label{intro}
Density estimation is essential in many aspects of astrophysics data interpretation: given a set of observed astronomical `points' e.g. stars, galaxies, quasars etc., we often require an estimate of the density of the underlying continuous distribution from which we could consider the observed sample as being drawn \citep[e.g.][]{2011A&A...531A.114F, 2015ApJ...805..121D}. This enables interpretation of the data and comparisons to theoretical models. However, like in many other scientific fields, noise might corrupt the observations, generically broadening the inferred distribution of our astronomical points and washing out important signals. Often we have a good understanding of the properties of this noise corruption allowing, in theory, for its removal and a more faithful representation of the underlying density, and in turn more robust inference on our underlying models. A flexible tool that can recover the underlying distribution in the presence of noise is therefore crucial for astrophysics study.

One concrete example of where such a tool is important is the estimation of properties of the Milky Way galaxy from catalogues of observations of individual stars. This is particularly timely given the results from the European Space Agency satellite, \emph{Gaia} \citep{Gaia}. \emph{Gaia} has provided an unprecedented and precise census of astrometric and photometric measurements for nearly $2$ billion stars. These measurements allow for the study of the Galaxy's structure, dynamics and evolution, the star formation history of the Galaxy, stellar physics and evolution, etc. \citep{gaiadr2_mapping,2019A&A...624L...1M,2021AJ....162..110K,antoja2023phase} The third data release, \emph{Gaia} DR3 \citep{Gaiadr3}, in June 2022, contained $\sim 32.2$ million stars brighter than $G_\mathrm{rvs} = 14$ with both astrometry and line-of-sight velocity measurements from \emph{Gaia}'s Radial Velocity Spectrometer (RVS) spectra \citep{2018A&A...616A...5C}. This full 6D position-velocity stellar sample allows for full phase-space density, or distribution function, characterization in the (extended) solar neighbourhood of the Milky Way and has revealed plenty of previously unknown kinematic substructures \citep[e.g.][]{gaiadr2_mapping, 2018MNRAS.478.3809S, 2021AJ....162..110K, 2023A&A...674A..37G}, most notably the \emph{Gaia} phase-space spiral \citep{Antoja_2018,antoja2023phase}. However, as previously highlighted, noise corrupts both the astrometric and spectroscopic observations from \emph{Gaia}, meaning that, although accurate characterization of the solar neighbourhood is possible without explicit consideration of the noise, extending analyses far beyond the solar neighbourhood becomes significantly harder. Furthermore, \emph{Gaia}'s noise model leads to significantly heteroscedastic uncertainties: for example, at a given Galactic location, we may have both bright giant stars, limited in number but with accurate astrometry, and faint dwarf stars, more numerous but with noisier astrometry. This heteroscedasticity makes analyses difficult but it is also advantageous as both smaller precise samples and larger noisier samples can accurately inform our Galaxy model.

Numerous approaches to density estimation have proved particularly useful for Milky Way analyses, both in the theoretical and observational setting. For example, kernel density estimation \citep[e.g.][for a recent example]{2024arXiv240811414L}, iterative partitions of the space \citep{2005MNRAS.356..872A, 2006MNRAS.373.1293S} and wavelet transforms \citep{2021A&A...651A.104P,2022A&A...666A..64R} have all found significant utility. However, typically these algorithms are not designed to deal with heteroscedastic noise in the observations \citep[although see][]{Delaigle2008}. A clear exception is the extreme deconvolution Gaussian mixture model \citep[XDGMM,][]{Bovy_2011} which utilises the analytic convolution of a Gaussian mixture model with Gaussian noise to obtain a noise-free distribution using the expectation-maximisation algorithm. XDGMM has proven useful across a range of astrophysical problems. For example, \citet{2018AJ....156..145A} use XDGMM to construct a colour--magnitude prior to improve the parallax estimation from \emph{Gaia}, whilst \citet{2015MNRAS.452.3124D} use XDGMM to generate probability density functions for quasar redshifts. However, despite its successes and relative simplicity, XDGMM training can be sensitive to the number of mixture components and their initialization, and fundamentally is limited to
\begin{inparaenum}
\item data with Gaussian uncertainties and
\item representing potentially significantly non-Gaussian features as a sum of Gaussians.
\end{inparaenum}

Here, we propose a novel machine-learning method that uses a normalizing flow model to estimate the distribution function. Normalizing flow models are deep-learning models that consist of a simple base distribution (typically a Gaussian) and a series of bijective transformations \citep{papamakarios2021normalizing}. These trainable transformations map the base distribution into the complex target distribution. Normalizing flow models have applications in astrophysics for complex density estimation or to speed up simulations. For example, \cite{ciuca2022unsupervised} used Neural Spline Flow \citep{Neural_Spline_Flows} to analyse the complex distribution of spectral space, \cite{lim2023mapping} mapped the dark matter distribution in the solar neighbourhood using \emph{Gaia} data and normalizing flows and \cite{wolf2023galacticflow} combined normalizing flows with galaxy formation simulations to speed up the simulation process. Their use on Milky Way phase-space data was emphasised by \cite{green2020deep} who demonstrated how equilibrium galaxy models could be constructed using normalizing flow models. \cite{Green_2023} tested this approach by recovering the gravitational potential from a toy model's phase space distribution function and \cite{kalda2023recovering} extended this method to successfully recover the gravitational potential of a simulated barred galaxy. Typically the emphasis in these works has been on the recovery of a smooth equilibrium model with little focus on the recovery of small-scale features.

In this work, we propose using a normalizing flow model for density estimation in a slightly different context to previous work in the astrophysics literature: for explicitly denoising 6D position-velocity data to obtain the noise-free underlying DF. We advocate that such a model is robust to the \emph{Gaia}-like heteroscedastic noise and demonstrate its power in recovering kinematic substructures through tests on artificially corrupted Milky Way observations. In Section~\ref{method_full} we detail our method along with describing the specifics of the normalizing flow architecture employed. In Section~\ref{data}, we describe how the model can be applied to \emph{Gaia}-like data and test the recovery of the noise-free distribution function from artificially corrupted \emph{Gaia} datasets. In Section~\ref{sec:discussion} we further discuss the hyperparameter choices for our setup, give a comparison with extreme deconvolution, highlight other useful normalizing flow architectures and detail potential applications of our approach. Our conclusions are given in Section~\ref{sec:conclusions}.

\section{Methodology}\label{method_full}
\subsection{Denoising using importance sampling}\label{method}

We begin by describing the problem before turning to our denoising methodology. We follow the notation from \cite{Density}. Consider a set of observed (multidimensional) data $w_i$ with ground truth values $v_i$ and additive noise $n_i$ such that:
\begin{equation}\label{additive_error}
    w_i = v_i + n_i.
\end{equation}
Suppose that the noise distributions $p_{n_i}(n_i)$ are known for all observed data. The observed distribution, $p(w_i)$, can then be related to the underlying `ground truth' distribution, $p(v)$, via a convolution
\begin{equation}
    p(w_i) = \int_{v} \mathrm{d}v\,p_{n_i}(w_i - v) p(v) .
    \label{}
\end{equation}
Our goal is to estimate $p(v)$ given the distribution of $w_i$ and known noise distributions, $p_{n_i}(n_i)$. Mathematically, this is a deconvolution. By insisting we know the noise distributions perfectly, this is a well-posed problem (particularly as we will only consider convolution kernels with everywhere positive Fourier transforms i.e. Gaussians). In practice, recovery of $p(v)$ can be a very sensitive procedure, particularly for recovering features well below the noise floor. 
\cite{CarollHall1988} discuss how for Gaussian noise, convergence of a non-parametric estimate of $p(v)$ is in the worst case logarithmic in the number of datapoints, and particularly slow for recovering `sharp' features.
From a probabilistic perspective, we are interested in the range of $p(v)$ consistent with the observations, which communicates this sensitivity and uncertainty. It may prove necessary to impose regularizing priors to restrict the range of $p(v)$ to `physically reasonable' distributions e.g. those without high-frequency structure. We will discuss this issue further later.
 
One approach to estimating $p(v)$ is kernel density estimation. This method is particularly efficient in Fourier space where deconvolution becomes division and is computationally practical when the noise distribution $p_{n_i}(n_i)$ is identical for all samples. There is significant discussion in the mathematical literature on the selection of the optimum kernel and convergence \citep[e.g.][]{Fan1991}. When the observational data have heteroscedastic noise distributions, i.e. $p_{n_i}(n_i)$ is different for each observation sample, alternate computationally efficient approaches have been employed. The extreme deconvolution (XD) method \citep{Bovy_2011} fits a Gaussian Mixture Model (GMM) for $p(v)$ convolved with a unique Gaussian noise distribution for each data point. In this case, the convolution is analytic and an expectation-maximization (EM) algorithm is used to optimize the objective function. Although the XDGMM model can approximate any distribution arbitrarily closely with enough mixture components, it lacks expressive power and can be hard to fit in practice (as we will demonstrate in Section~\ref{sec:xdgmm}).

Normalizing flow models are attractive alternatives for modelling arbitrary density distributions. These models map a simple underlying density distribution (for example, a Gaussian) into a complex target density distribution through a series of non-linear transformations. We will describe how this is practically implemented below. For understanding our approach, it is only necessary at this stage to consider a normalizing flow model as some flexible way to model a probability density function, $P_\mathrm{NF}(v|\Theta)$, that depends on a set of hyperparameters $\Theta$. \cite{Density} provide a general framework for applying normalizing flows to density deconvolution, which takes a variational approach by estimating evidence lower bound (ELBO) using Variational Auto-Encoders \citep{vae}. This framework allows for density deconvolution for arbitrary noise distributions. We adopt a different approach that estimates $p(w_i)$ through importance sampling. We assume that we are able to efficiently sample $v$ from the distribution $p_{n_i}(w_i-v)$. This makes our approach more limited than that of \cite{Density} but in practice the noise distribution of most observational data is from easy to sample distributions e.g. Gaussian, Poisson. The likelihood of the observation $w_i$ is then computed as
\begin{equation}
   p(w_i) = \mathbf{E}_{v\sim p_{n_i}(w_i-v)} [ P_\mathrm{NF}  (v|\Theta) ]= \sum_{j=1}^{K_i} P_\mathrm{NF}(\hat{v}_{i,j}|\Theta).
\end{equation}
$\hat{v}_{i,j}$ are the z-scored(transformed to have mean zero and unit standard deviation samples generated from the $i$-th noise distribution, $p_{n_i}(n_i)$. $K_i$ here is the (variable) number of samples per star. The middle term $\mathbf{E}_{v\sim p_{n_i}(w_i-v)} [ P_\mathrm{NF}  (v|\Theta) ]$ is the expectation value of the probability from the normalising flow model with respect to the uncertainty distribution. For our considered use case (the \emph{Gaia} data), the noise distribution is simply Gaussian, $n_i\sim \mathcal{N}(n_i|0,\Sigma_i)$ for covariance matrix $\Sigma_i$, making the generation of samples $v$ simple and efficient. The total log-likelihood loss $\mathcal{L}$ is given by
\begin{equation} \label{eq:loss}
    \mathcal{L}  =  \sum_{i=1}^N \ln \frac{1}{K_i} \sum_{j=1}^{K_i} P_\mathrm{NF}(\hat{v}_{i,j}|\Theta).
\end{equation}
Given the flexibility of the model, $P_\mathrm{NF}(v)$, and the potentially `large' uncertainties in the data, we have found it advantageous to split model training into two stages: 1. a pre-train loop where the uncertainties are completely ignored, and 2. a full training loop initialized using the pre-trained model where the uncertainties are incorporated using our importance sampling scheme. A further advantage of this two-stage procedure is that the first stage is computationally cheaper than the second stage (by a factor of $\sim\langle K_i\rangle$ so allows for rapid testing. The log-likelihood loss of the pre-train stage $\mathcal{L}_\mathrm{pre}$ is simply given by
\begin{equation}
    \mathcal{L}_\mathrm{pre}  =  \sum_{i=1}^N \ln P_\mathrm{NF}(\hat{v}_{i}|\Theta).
    \label{eq:loss_pretrain}
\end{equation}

\subsection{Choice of normalizing flow architecture}\label{architecture}

We have described a generic approach to denoising a point cloud contaminated by Gaussian noise using a general flexible `ground truth' probability density function, $P_\mathrm{NF}$. As already suggested, normalising flow models are ideal tools for this flexible representation, as they transform Gaussian distributions into arbitrary target distributions. A normalising flow architecture is advantageous due to its expressive power and computational efficiency. There are multiple possible choices of architecture detailed in the literature. Normalising flow models can be classified into several categories based on their bijection transformation. The most straightforward affine transformation is used in models such as Nice \citep{Nice}, Masked Autoregressive Flow \citep[MAF]{papamakarios2018maf} and Glow \citep{kingma2018glow}. Some models instead use spline-based transformations such as the linear and quadratic spline models \citep{Neural_Importance_Sampling} and the rational quadratic spline model \citep{Neural_Spline_Flows}. There are also transformers that use iterative Gaussianization \citep{meng2020gaussianizationflows}, Bernstein polynomials \citep{sick2020bernstein-polynomial}, or even replace the chain of discrete transformers with a continuous ODE transformer \citep{grathwohl2018ffjordfreeformcontinuousdynamics}. We must select an architecture from this wide array of options that is most suitable.
Our main requirement is that the architecture needs to be robust and efficient. Therefore, we tend to choose the simplest architecture that captures enough details in the pre-train stage, i.e. the training loop that ignores uncertainties, whilst balancing between expressive power and robustness/model efficiency.

A further consideration is the dimensionality of our dataset. The \emph{Gaia} astrometric plus spectroscopic data is only six-dimensional. This is relatively low compared to other tasks that normalizing flows have been employed for: for example \citet{ciuca2022unsupervised} use normalizing flow models on $2405$-dimensional spectral data. With our low dimensionality, we employ an autoregressive architecture \citep{papamakarios2021normalizing} over dimensions. Furthermore, we choose the element-wise transformation to be the rational quadratic spline transformation \citep{Neural_Spline_Flows}. Rational quadratic spline transformers generate the bijective map using splines. The splines separate the fixed input/output interval into several bins by assigning a series of monotonically increasing coordinates called knots. Given the value and derivative at the knots, each bin is defined by a monotonic rational quadratic function (i.e. the quotient of two rational polynomials). The coordinates of the knots and the corresponding derivatives are then the trainable parameters. Rational quadratic splines have the potential to be more expressive than linear or quadratic splines, so can capture small-scale features while remaining efficient in training.
We tested the LU factorization technique of \cite{kingma2018glow}, but found that it does not improve the performance for our tests so it is not included in our architecture.

For the rational quadratic spline model architecture, the hyperparameters are the layer number $L$, the hidden unit $h$, and the number of bins $B$ of the spline function. We set the flow model to have six layers, $L = 6$. Each rational quadratic spline layer has two residual blocks that contain two dense layers with hidden features $h = 128$ and the bin number $B = 32$. We also have an additional hyperparameter that controls the fidelity of the denoising, the number of samples $K_{i}$. We will discuss the choice of this parameter more fully later.
Rational quadratic spline flow models require the specification of a boundary (or a transformation of an infinite to finite domain) for both input and output. Our input data is always z-scored and a soft clipping is applied to restrict the input scaled coordinate to $|v|>5$.
We use the Adam optimiser \citep{2014arXiv1412.6980K} with the loss function given in equation~\eqref{eq:loss_pretrain} and set the initial learning rate, $r$, to $0.001$. After $\texttt{epoch}_\texttt{max pre-train}=100$ pre-train epochs, the model is further trained for $\texttt{epoch}_\texttt{max}$ denoise epochs using the loss function in equation~\eqref{eq:loss} with an exponent learning schedule with $\gamma = 0.99$. The batch size affects the training significantly. Due to hardware limitations, as we increase the number of denoising samples $K_i$, we reduce the batch sizes from a maximum of$1024$ to a minimum of $256$. The \textsc{Zuko} \citep{2023zndo...7625672R} and \textsc{Pytorch} \citep{2019arXiv191201703P} libraries are used to construct the normalising flow model. The algorithm is summarised in Algorithm~\ref{alg}.

\begin{algorithm}
\begin{algorithmic}
\State $w^\mathrm{obs}_{i}, \sigma^\mathrm{obs}_i \gets \mathrm{Gaia\,Data} $ \Comment{Select stars within cylindrical region}

\While{$\texttt{epoch} <\texttt{epoch}_\texttt{max pre-train}$} \Comment{pre-train section}

\State $\{w^\mathrm{gc}_{i}\}\gets \{w^\mathrm{obs}_{i}\}$ \Comment{Galactocentric Coordinate}
\State $\{\hat{w}_{i}\} \gets \{w^\mathrm{gc}_{i}\}$ \Comment{Standard Scalar}
\State $\mathcal{L}_\mathrm{pre}  =  \sum_{i=1}^N \ln P_\mathrm{NF}(\hat{w}_{i}|\Theta) $
\State $\Theta \gets r \nabla \mathcal{L}_\mathrm{pre}$ \Comment{Back propagation}
\State $r \gets \delta \mathcal{L}_\mathrm{pre} $ \Comment{Update learning rate}

\EndWhile

\While{$\texttt{epoch} <\texttt{epoch}_\texttt{max}$} \Comment{de-noise section}
\State $\{v^\mathrm{obs}_{i,j}\}_1^{K_i} \gets  \mathcal{N}(w^\mathrm{obs}_{i}, \sigma^\mathrm{obs}_i) $ \Comment{ Sampling}
\State $\{v^\mathrm{gc}_{i,j}\}_1^{K_i}\gets \{v^\mathrm{obs}_{i,j}\}_1^{K_i}$ \Comment{Galactocentric Coordinate}
\State $\{\hat{v}_{i,j}\}_1^{K_i} \gets \{v^\mathrm{gc}_{i,j}\}_1^{K_i}$ \Comment{Standard Scalar}
\State $\mathcal{L}  =  \sum_{i=1}^N \ln \frac{1}{K_i} \sum_{j=1}^{K_i} P_\mathrm{NF}(\hat{v}_{i,j}|\Theta) $
\State $\Theta \gets r \nabla \mathcal{L}$ \Comment{Back propagation}
\State $r \gets \delta \mathcal{L} $ \Comment{Update learning rate}
\EndWhile
\end{algorithmic}
\caption{Denoising algorithm with a pre-train stage.}
\label{alg}
\end{algorithm}

\section{Application to the \emph{Gaia} dataset}\label{data}

With our general procedure for fitting a normalising flow model to a noisy point cloud described, we now address the specifics of handling our input data, the \emph{Gaia} data. Although our goal is to apply the denoising procedure in the regime where noise significantly impacts the observations, we will perform tests of our procedure by injecting artificial noise into the local high-quality \emph{Gaia} data.

\subsection{Gaia data and preprocessing}
We use the \emph{Gaia} DR3 data with radial velocities from \emph{Gaia} RVS selected with a quality cut of
\begin{equation}
    \texttt{parallax}> 0 \quad \mathrm{and}\quad 
    \texttt{error\_over\_parallax} > 0.5.
\end{equation}
These cuts will introduce biases in recovering the genuine Galactic distribution function. However, in this work, the sampled Gaia data only serves as an example ground truth test set with realistic dynamical features with which we can test the flow models. In realistic applications, more care over data selection would be required to not bias any inference.
Our observable data $w^\mathrm{obs}_{i}$ are on-sky positions $(\alpha, \delta)$, parallax $\varpi$ and the proper motion vector $(\mu_\alpha\cos\delta, \mu_\delta)$ i.e. a 6D vector per star. The \emph{Gaia} DR3 catalogue reports the Gaussian uncertainties, $\sigma^\mathrm{obs}_i$, for all of these measured quantities. For simplicity, we ignore the negligible errors in on-sky positions and set the correlations in the covariance matrices of the observables to zero. The 6D \emph{Gaia} observables are then transformed into Galactocentric coordinate using \textsc{Astropy} \citep{2022ApJ...935..167A}, with solar coordinate $(R_{\odot},\phi_{\odot},Z_{\odot}) = (8.122 \, \mathrm{kpc},180 \,\mathrm{deg},0.0208 \, \mathrm{kpc})$ and Cartesian velocity $(U,V,W) = (12.9,245.6,7.78) \, \mathrm{km/s}$. For our tests, we further select stars in a region that is centred on the Sun, with Galactocentric cylindrical radius $R = R_{\odot} \pm 0.1 \, \mathrm{kpc}$, azimuthal angle $\phi = 180 \pm 0.1 \,\mathrm{deg}$ and Galactic height $z = \pm 2.0 \,\mathrm{kpc}$. This region covers large range of vertical direction for 3D feature like phase spiral.  The total number of stars within this region is $761,500$.

We first generate $K_i$ samples $\{v^\mathrm{obs}_{i,j}\}_1^{K_i}$ from the Gaussian distributions $\mathcal{N}(w^\mathrm{obs}_{i}, \sigma^\mathrm{obs}_{i})$ for each data sample $w_i$. $K_i$ could be different for each star, e.g. a function of the uncertainty $K_i = f(n_i)$ such that fewer samples are used for stars with more precise measurements (which presents complications in vectorizing calculations). However, in our tests, we use the same number of samples for every star, i.e. $K_i=K$.
We then apply a coordinate transformation to transform to Galactocentric cylindrical position and velocity coordinates $\{v^\mathrm{gc}_{i,j}\}_1^K$. We train our model in the Galactocentric coordinate to examine the denoising ability over features in this coordinate. In this way, the sampling and transformation procedures are separate from the flow model. An alternative procedure is to explicitly train the flow model on observable coordinates and embed the Galactocentric transformation inside the flow model. However, this procedure is more complicated as we would require special treatment of the periodic angle variables in the observable space. Therefore, for simplicity, we keep the transformation separate and train the model in the physically-interesting Galactocentric space.
Finally, the standard scalar transformation is applied to those coordinates such that
\begin{equation}
    \hat{v}_{i,j} = \frac{v^\mathrm{gc}_{i,j} - \Bar{v}^\mathrm{gc}}{\Bar{s}^\mathrm{gc}},
\end{equation}
where $\Bar{v}^\mathrm{gc}$ and $\Bar{s}^\mathrm{gc}$ are the means and standard deviation of all selected training data.
Note that for each epoch, a new sample $\hat{v}_{i,j}$ is drawn instead of reusing the same samples. By drawing new samples, the training averages over different realisations of the `true' distribution for each datum at each epoch meaning we avoid fitting features that might arise in a fixed sample and we can perhaps use a lower number of samples per epoch to take advantage of this averaging. For a non-physically generated sample, e.g. a sample with negative parallax, the probability, $p(v)$, is set to zero.

\subsection{Artifically corrupted \emph{Gaia} data test sets}\label{sec:corruption}

To test the denoising ability of the described method, we deliberately modify the local high-quality \emph{Gaia} data by increasing the uncertainties and then re-sampling with this mock error to generate a noisy mock data sample. These mock data represent how the noise distorts kinematic features and is used to train the model. The original \emph{Gaia} data serves as a ground truth distribution to test the ability of the denoising model to reconstruct those distorted kinematic substructures.

For the experiment, we split the Gaia data into train and test data with training proportion $\eta$. The mock errors $\sigma^\mathrm{mock}$ are generated by amplifying the reported \emph{Gaia} errors $\sigma^\mathrm{obs}$ of the train data by a factor of $q$:
\begin{equation}
    \sigma^\mathrm{mock} = q\sigma^\mathrm{obs}.
\end{equation}
Mock data are then generated by adding linear noise to the training data following equation~\eqref{additive_error}. We use a simple uncorrelated Gaussian distribution for each dimension:
\begin{equation}
    w^\mathrm{mock} = w^\mathrm{obs} + \mathcal{N}(0,\sigma^\mathrm{mock}).
\end{equation}
In our 6D model tests, we choose $q = 5$ and $q = 10$ which means convolving the solar neighbourhood data with $5$ times and $10$ times the reported \emph{Gaia} uncertainty respectively. This inflated mock error covers the uncertainty range of the entire \emph{Gaia} RVS data. Therefore, it is a good representation of regions outside the solar neighbourhood.

We also provide two different normalising flow benchmark results:
\begin{enumerate}
    \item \emph{No deconvolution}: 
    we train the flow with the mock data \emph{directly} without generating any samples from the error distributions. This is equivalent to our pre-train stage during training (i.e. using the loss function in equation~\eqref{eq:loss_pretrain}). This setup represents a benchmark with noisy data ignoring the error.
    \item \emph{Ground truth}:
    we train the flow with the \emph{Gaia} data \emph{directly} without any noise contamination (i.e. $q=0$) and treat this data as entirely noise-free. This setup represents the ``ground-truth'' underlying distribution which our denoising procedure aims to recover i.e. it is the best case scenario for the denoising process.
\end{enumerate}
We train all models via the two-stage procedure described in Section~\ref{method}, always maximising the mean log-likelihood. After training, we compute the test loss by evaluating the mean log-likelihood for the normalizing flow model on a held-out test set of the original uncorrupted \emph{Gaia} data. We use the same test set to get a meaningful comparison of loss values from different models. The test set is a fixed 30\% of the Gaia data. We compare this test loss with the benchmarks to evaluate the denoising ability. 

\subsection{Fiducial setup results}\label{6dresult}

\begin{figure*}
  \centering
\begin{subfigure}[b]{.5\textwidth}
  \centering\includegraphics[width=\textwidth]{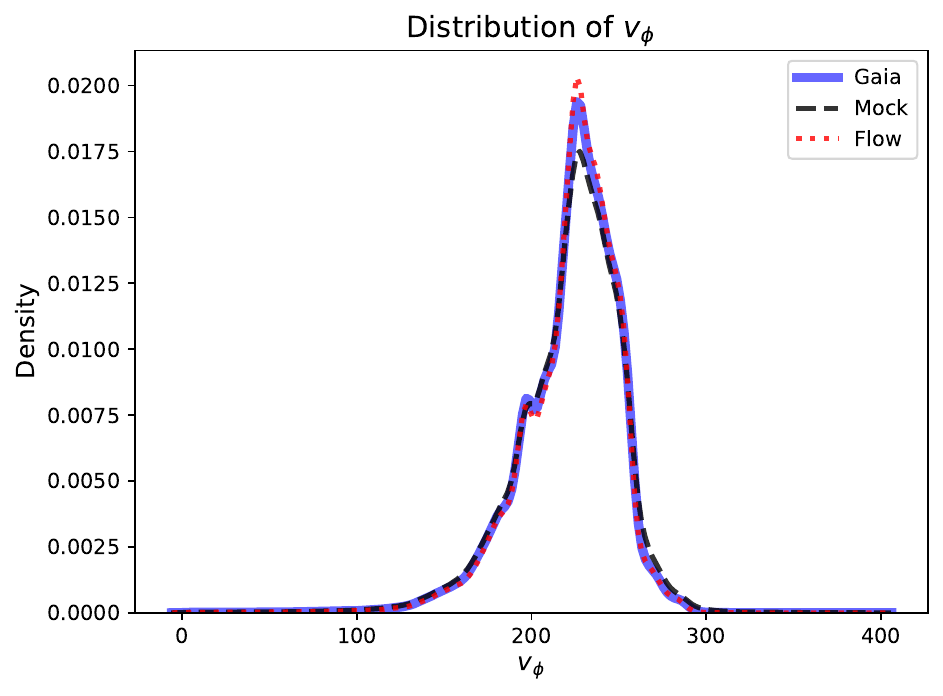}
\label{fig:6d_vphi_1d_amp5}
\caption{$q=5$}
\end{subfigure}%
\begin{subfigure}[b]{.5\textwidth}
  \centering
  \includegraphics[width=\textwidth]{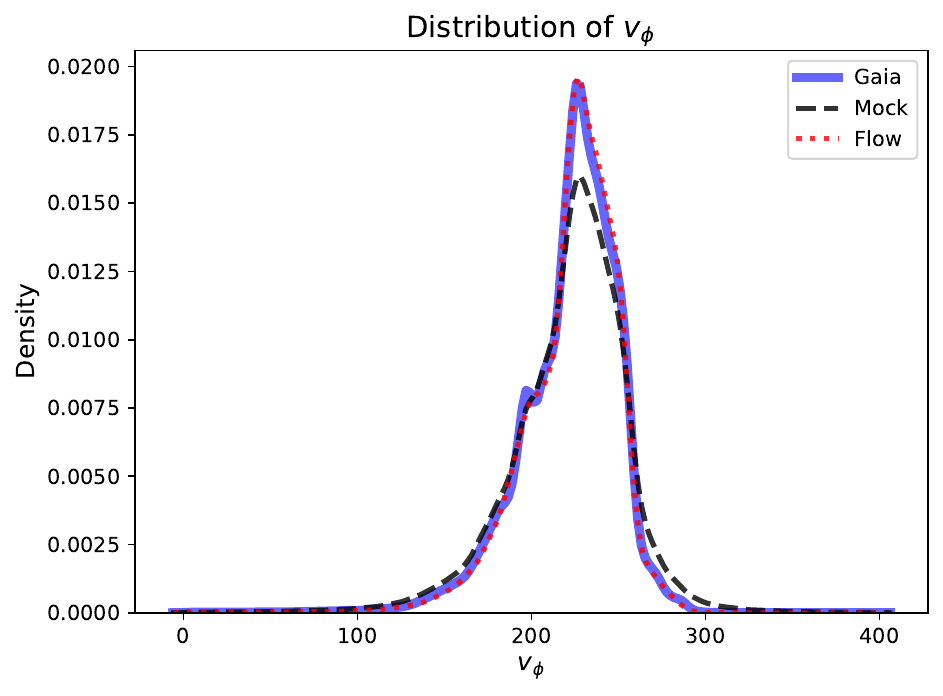}
\label{fig:6d_vphi_1d_amp10}
\caption{$q=10$}
\end{subfigure}
\caption{1d $v_\phi$ distribution for models trained using the $q=5$ (left) and $q=10$ (right) training sets. For the $q=5$ setup, the flow model successfully recovers the Hercules Stream feature at $200\,$km/s, whilst for $q=10$ the Hercules feature is not as cleanly resolved.}
\label{fig:6d_vphi_1d}
\end{figure*}
We begin by showing the results 
% for the $q=5$ mock data sets 
using a `fiducial setup' with $K=128$ and $\eta=0.7$ training data fraction ($536109$ stars). Although we will discuss the quality of the denoising procedure quantitatively below, we will first focus on the success of the procedure in resolving known features in the solar neighbourhood sample. When plotting the fitted models to compare with the data, we always draw a model sample of equal size to the data sample.
First, we show the 1d $v_\phi$ distributions for this setup in Fig.~\ref{fig:6d_vphi_1d}. We see that the flow model trained on the $q=5$ data successfully recovers the 1d $v_\phi$ distribution from the mock data distribution. There is minor overfitting at the peak, but the model also recovers the Hercules Stream, the bump at $v_{\phi} = 200\,\mathrm{km\,s}^{-1}$. Although the $q=10$ flow model broadly recovers the distribution, it fails to recover the Hercules bump. The full 6d corner plots are shown in the appendix, with Fig.~\ref{fig:corner_amp5} and Fig.~\ref{fig:corner_amp10}, which indicates the denoising ability of flow model.

\begin{figure*}
    \centering
    \includegraphics[width=\textwidth]{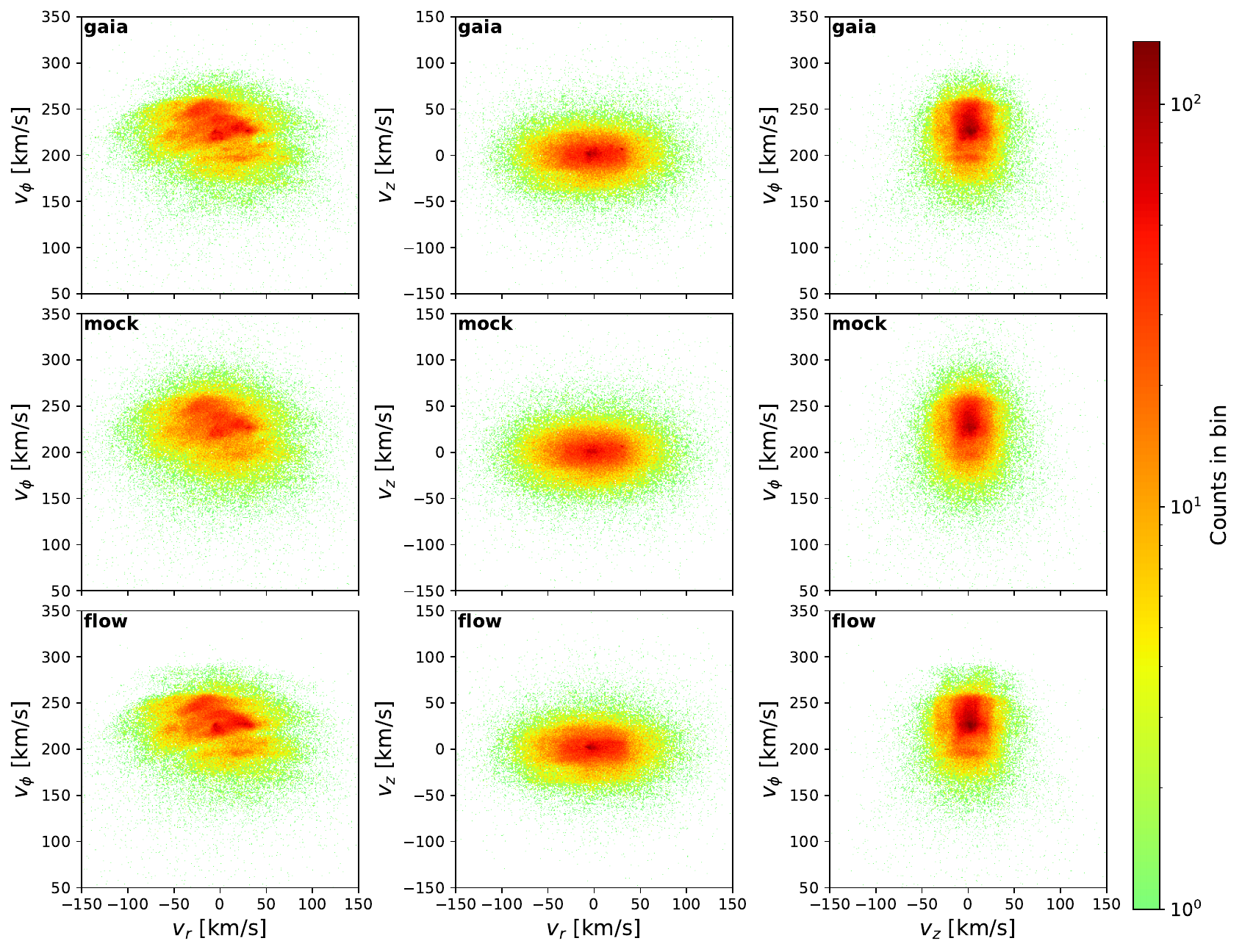}
    \caption{The velocity distribution within $z = \pm 0.05\,\mathrm{kpc}$ for the considered `ground-truth' \emph{Gaia} training dataset of $536109$ stars ($\eta=0.7$, top row), a mock training dataset generated by scattering the training dataset by $q=5$ times their uncertainties (middle row) and samples from the fitted normalizing flow model using $K=128$ samples per star from the uncertainty distributions. Note the similarity between the top and bottom rows, and how the model (bottom row) has managed to remove the noise introduced in the mock dataset (middle row). We note that
    \begin{inparaenum}
    \protect\item in the left column, the flow model clearly enhances the moving groups, e.g. the Hyades at the centre of the velocity peak, and recovers the boundaries between the Hercules stream arches at $v_{\phi} = 200$ km/s. The model also recovers a cleaner boundary at $v_{\phi} = 280$ km/s compared to the noisy mock data;
    \protect\item in the middle column, the flow model recovers the `shell' at  $v_z$ = -30 km/s;
    \protect\item in the right column, the flow model separates the arch features at $v_{\phi}$ = 200 km/s and constrains the distribution within the central region.
    \end{inparaenum}
    }
    \label{fig:6d_v_0.05_amp5}
\end{figure*}

To further check the ability of the model to reproduce fine disc structure e.g. the Hercules stream, we consider the velocity diagrams within $z = \pm 0.05$ kpc for the $q=5$ case in Fig.~ \ref{fig:6d_v_0.05_amp5}. We show the mock \emph{Gaia} training data ($\eta = 0.7$, $536109$ stars) and compare with samples from the flow model. We generate the same number of samples from the flow model as in the full mock training data before restricting the $|z|$ range. This reduced sample is only 2.5 per cent of the full dataset. Fig.~ \ref{fig:6d_v_0.05_amp5} shows that the flow model for the $q=5$ training dataset recovers the substructures mentioned in \cite{Antoja_2018} despite the added noise. For example, the boundary of the thin arches of Hercules streams in the $v_r$ vs. $v_{\phi}$ plot and the shells related to the phase-space spiral pattern in $v_z$ vs. $v_{\phi}$ are recovered. We provide results for the noisier $q=10$ test in Appendix~\ref{amp10}. As well as a residual plot between Gaia data and flow sample in Appendix~\ref{residual}.

\begin{figure*}
    \centering
    \includegraphics[width=0.9\textwidth]{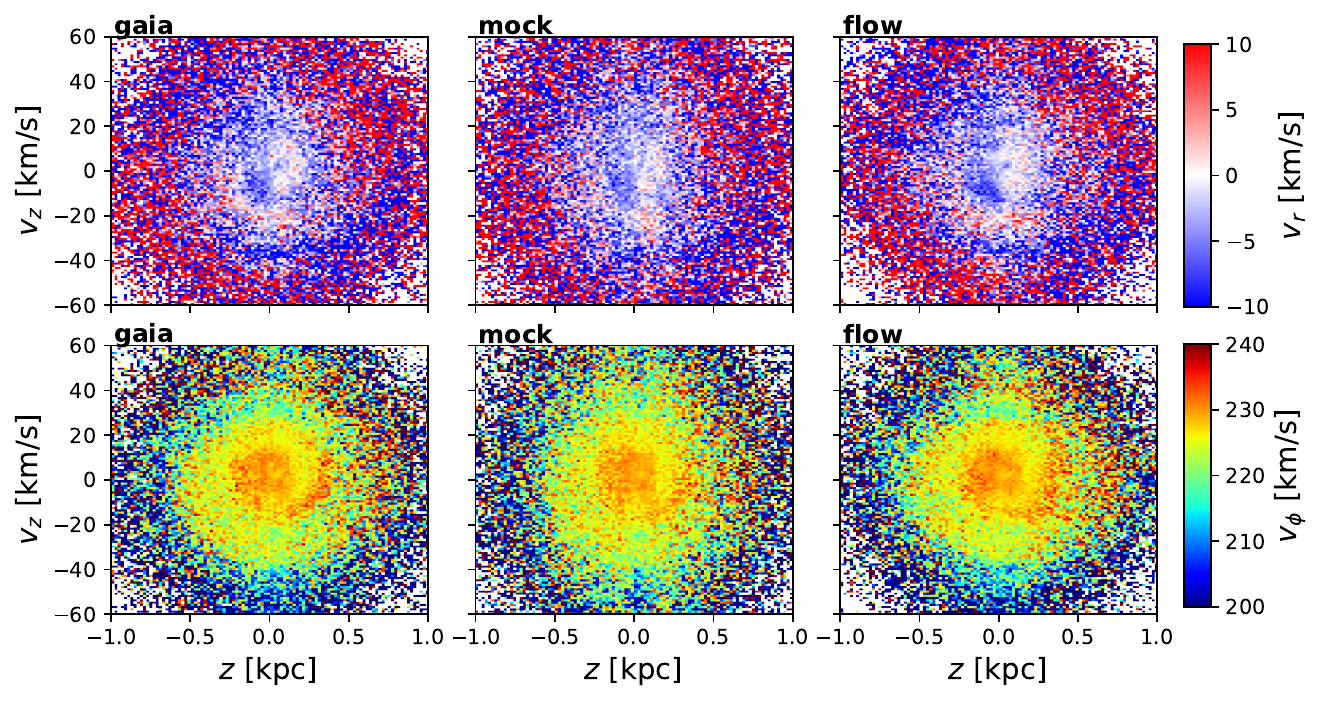}
    \caption{Phase spiral recovered from the $q=5$ mock dataset of $536109$ stars ($\eta=0.7$). We show two projections: $(z,v_z,v_r)$ top row and $(z,v_z,v_\phi)$ bottom row. The left column shows the `ground-truth' \emph{Gaia} dataset, the middle column the mock dataset and the right column samples from the normalizing flow model of the denoised distribution. The normalizing flow model enhances the spiral from the mock data and corrects the distorted feature in the plane ($z=0$).}
    \label{fig:6d_spiral_amp5}
\end{figure*}

Our final qualitative check of the model is its reproduction of the phase spiral. This is a significantly harder feature to capture due to both its morphology and that it is predominantly a 3d structure i.e. it was first observed clearly in 2d vertical phase space $z$ vs. $v_z$ coloured by $v_\phi$ (although it can be seen more weakly purely as an overdensity in the 2d vertical phase space). We show projections of $(v_z,z,v_{\phi})$ and $(v_z,z,v_r)$ for the $q=5$ mock data configuration in Fig.~\ref{fig:6d_spiral_amp5}, where the colour bar shows the median value of the represented bin. Recall we have selected Gaia data with $R = R_{\odot} \pm 0.1 \, \mathrm{kpc}$, azimuthal angle $\phi = 180\pm 0.1 \,\mathrm{deg}$ and Galactic height $z = \pm 2.0 \,\mathrm{kpc}$, which only cover a relatively small region on the disc and hence the phase spiral is noisier than in other studies utilising a broader sample. The model for the $q=5$ training set successfully removes the noise in the Galactic plane ($z=0$) and enhances the spiral feature. These results show the flow model has the capability to recover complex 3d features from the distorted data. Again, in Appendix~\ref{amp10}, we show the equivalent results for the $q=10$ training dataset. 

\subsection{Model variants}
We have shown how our fiducial choice of $K=128$ successfully reproduces structures in the solar neighbourhood test set, at least qualitatively, when using $536,109$ stars ($\eta=0.7$). We will now quantitatively investigate how the performance varies with different numbers of samples $K$ and different training set sizes.

\subsubsection{Varying the number of samples per star, $K$}\label{sec::varying_K}
The most significant hyperparameter for denoising is $K_i$, the Monte-Carlo sample size per star for the deconvolution. The training time scales as $K$, so choosing an appropriate $K$ is crucial for balancing denoising ability and computational cost.

\begin{figure}
    \centering
    \includegraphics[width=0.481\textwidth]{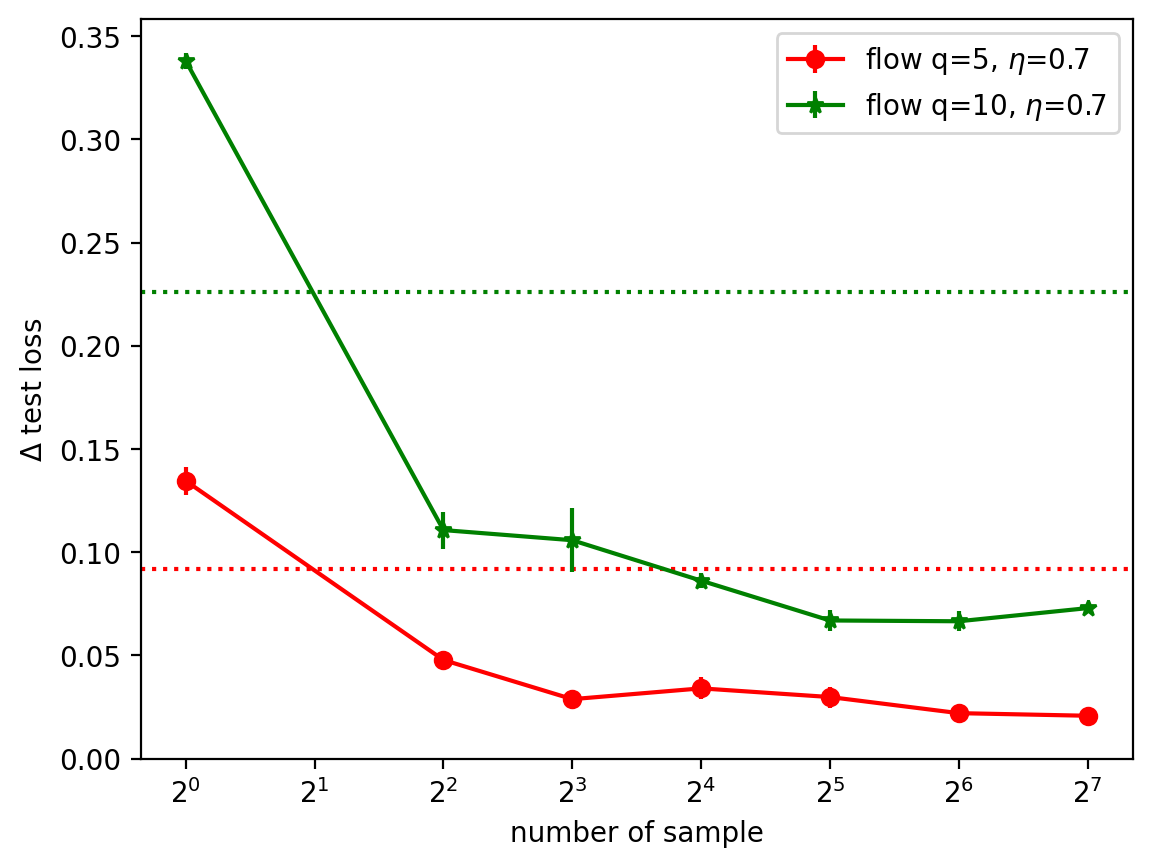}
    \caption{Test loss per star for the models trained on the $q=5$ (red dots) and $q=10$ (green crosses) datasets of $536109$ stars ($\eta=0.7$). The test loss is computed on the held-out test set and is reported relative to the best-case scenario loss computed using the `ground truth' dataset with no noise. The dotted lines show the loss for models with no deconvolution.}
    \label{fig:test_loss_amp10}
\end{figure}

Table~\ref{tab:6dloss} and Fig.~\ref{fig:test_loss_amp10} show the test loss  when $K$ is varied from $K=1$ up to $K=128$. We report the average test loss relative to the test loss for the ground-truth noiseless distribution and its standard deviation over five different runs per $K$ choice. The test loss decreases as the number of samples increases, indicating an improvement in the denoising ability. We see that for both $q=5$ and $q=10$, we fail to precisely reproduce the ground truth model even for large $K$. There is no further improvement in the test loss for $K > 128$. However, despite the failure to exactly converge to the `ground truth' loss (for the reasons discussed in Section~\ref{method_full}), Fig.~\ref{fig:6d_v_0.05_amp5} shows that the bulk of features are still captured for this case. The test loss shows that even with a small number of samples per star, the denoised model is still better than completely ignoring the uncertainty.

\begin{table}
    \caption{Test loss per star for 6d models trained using the $q=5$ and $q=10$ datasets of $536109$ stars ($\eta=0.7$) and varying $K$, the number of samples per star used in the denoising.}
    \centering
    \begin{tabular}{|c|c|c|}
        &\multicolumn{2}{c}{Test loss}\\
        Setup & $q=5$ dataset & $q=10$ dataset\\
        \hline
        $K=1$  & 13.16156 $\pm$ 0.00681 & 13.33386\\
        $K=4$  & 13.07911 $\pm$ 0.00399 & 13.15877\\
        $K=8$  & 13.05419 $\pm$ 0.00277 & 13.13957\\
        $K=16$  & 13.048406 $\pm$ 0.00539 & 13.10281\\
        $K=32$  & 13.04421 $\pm$ 0.00508 & 13.08481\\
        $K=64$  & 13.03640 $\pm$ 0.00266 & 13.08559\\
        $K=128$  & 13.035136 $\pm$ 0.00369 & 13.08729\\
        \\
        No deconvolution  & 13.10600 &  13.24022\\
        \\
        Ground truth & 13.01441 & 13.01847\\
    \end{tabular}
    \label{tab:6dloss}
\end{table}

\subsubsection{Varying the training data size, $\eta$}\label{sec::varying_data_test}
We test the robustness of the flow model with smaller datasets by changing the training proportion $\eta$ of the \emph{Gaia} data. The following test uses three proportions: $\eta = 0.7$ (536109 stars in the training data, default as used above), $\eta = 0.3$ (229761 stars in the training data), and $\eta = 0.1$ (76587 stars in the training data). Note that changing the split proportion also changes the test set, and hence changes the test loss of the \emph{No deconvolution} benchmarks. We use $q=5$ for these tests.

\begin{figure}
    \centering
    \includegraphics[width=0.481\textwidth]{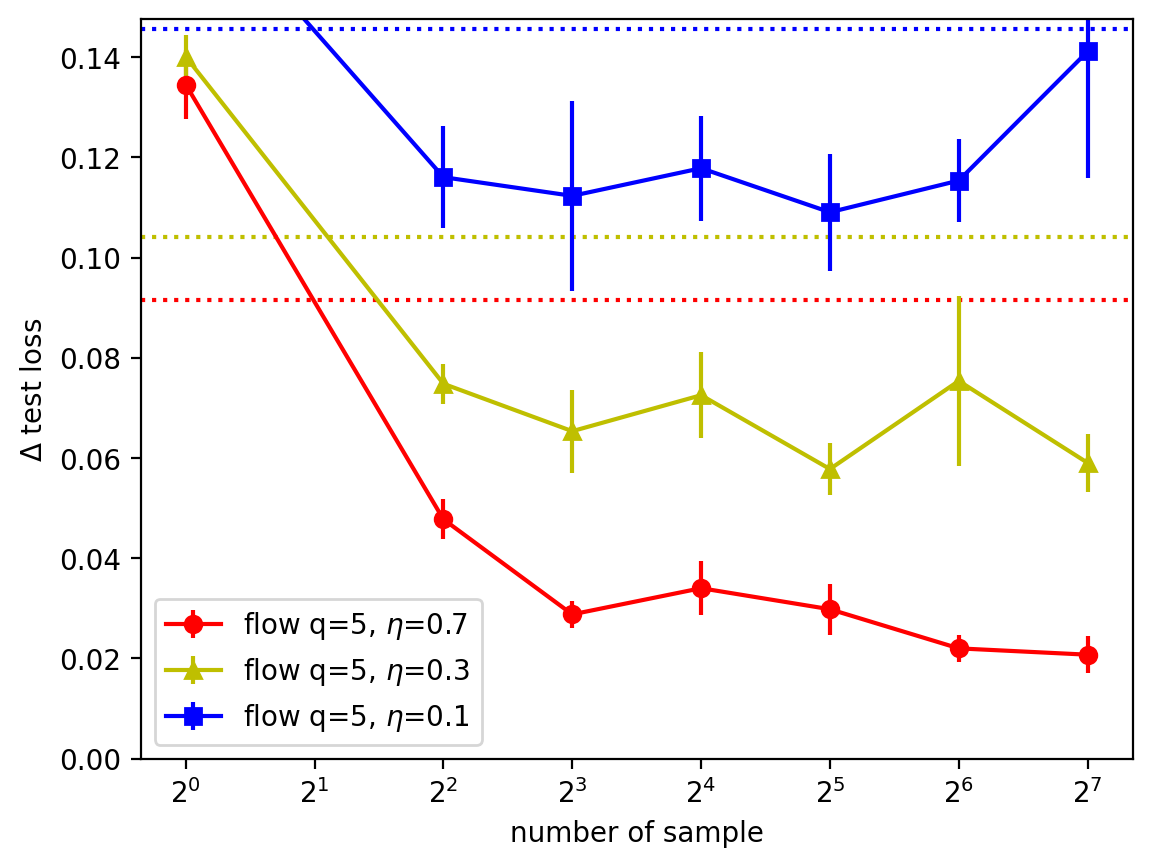}
    \caption{Test loss per star of 6D model for different data sizes (as shown by the different symbols and colours) trained using the $q=5$ dataset. Like Fig.~\ref{fig:test_loss_amp10}, the test losses are plotted relative to the best-case scenario computed using the `ground-truth' dataset with no noise. The dotted lines show results without any denoising. }
    \label{fig:test_loss_amp5}
\end{figure}

As shown in Table \ref{tab:6dlossratio} and Fig.~\ref{fig:test_loss_amp5} , the normalizing flow models successfully denoise the smaller datasets. However, we observe weaker improvement of the results with increasing $K$ than in our baseline setup. When using $\eta = 0.3$ (i.e. $229,761$ stars) of the original dataset, the flow model reaches a test loss plateau for $K>8$. When using $\eta = 0.1$ (i.e. $76,587$ stars), the plateau is reached at $K\geq4$. Beyond these $K$ values, the Poisson error arising from sampling the uncertainty distributions is smaller than the Poisson error arising from the sampling of the underlying distribution function. In the next section, we discuss how reducing the data size further ($\eta=0.01$) leads to model artefacts.

\begin{figure*}
    \centering
    \includegraphics[width=0.9\textwidth]{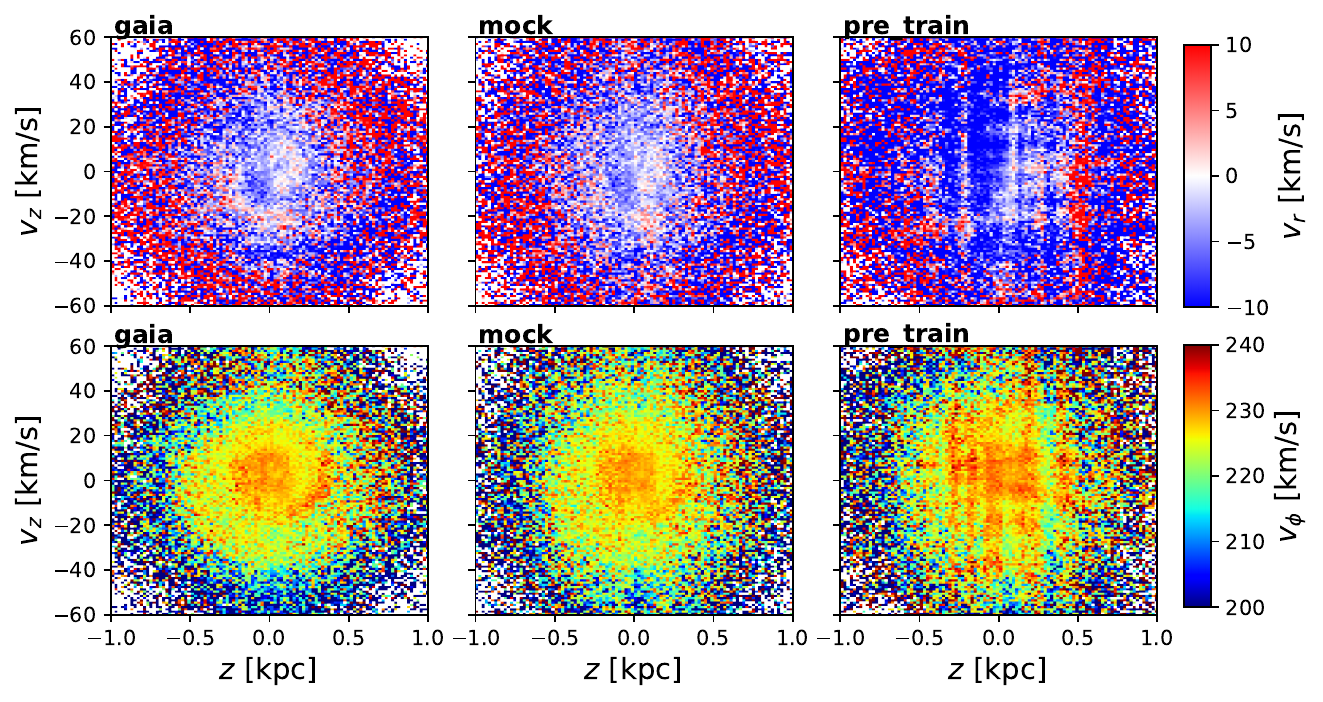}
    \caption{Phase-space spiral plot when  using only 1\% of the training data ($7659$ stars). For better comparison between \ref{fig:6d_spiral_amp5}, the first and second column shows the spiral plot \emph{Gaia} data and the mock data of $\eta = 0.7$, 536109 stars in total. And the third column shows the 536109 samples that generate from the model train with only 7659 stars.  The fitted flow model exhibits gridded artefact features due the limited data not constraining the overly flexible model. These features are accentuated by the oversampling.}
    \label{fig:0.01-spiral}
\end{figure*}

\begin{figure*}
    \centering
    \includegraphics[width=\textwidth]{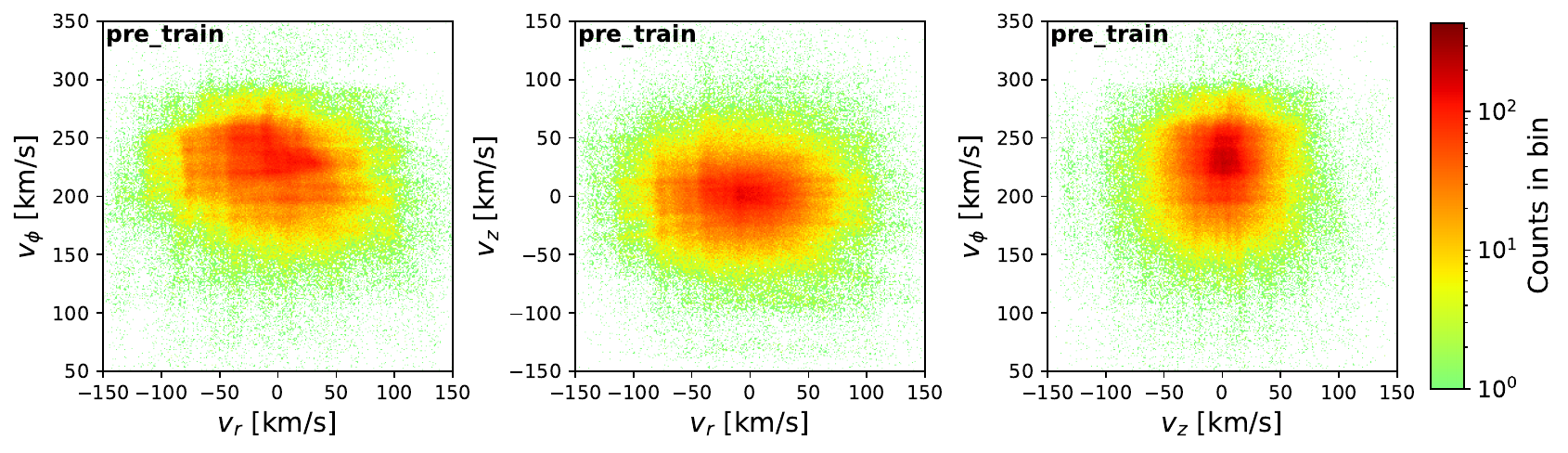}
    \caption{Velocity distribution of the flow model using only 1\% of the training data ($7659$ stars). Similarly, the first and second column shows the spiral plot \emph{Gaia} data and the mock data of $\eta = 0.7$, 536109 stars in total. And the third column shows the $536109$ samples that generate from the model train with only 7659 stars. Gridded artefact features as Fig~\ref{fig:0.01-spiral}.}
    \label{fig:0.01-velocity}
\end{figure*}

\begin{table}
    \caption{Test loss per star for 6d model using different training proportions, $\eta$, of the full dataset for training ($70$, $30$ and $10$ per cent of the full $761300$ star dataset). Mock data are generated using $q=5$.}
    \centering
    \begin{tabular}{|c|ccc}
        &\multicolumn{3}{c}{Test loss}\\
        Setup  & $\eta = 70\,\%$ & $\eta = 30\,\%$ & $\eta = 10\,\%$ \\
        \hline
        $K=1$  & 13.14892 & 13.15448 & 13.18886\\
         & $\pm$ 0.00681 & $\pm$ 0.00442 & $\pm$ 0.00715 \\
        $K=4$  & 13.062154 & 13.089242 & 13.130435 \\
         & $\pm$ 0.00399 &  $\pm$ 0.00374 & $\pm$ 0.01019\\
        $K=8$  & 13.05419 & 13.07974  & 13.127017 \\
        & $\pm$ 0.00276 & $\pm$ 0.00858 & $\pm$ 0.01921\\
        $K=16$  & 13.048406  & 13.08680  & 13.132312 \\
        & $\pm$ 0.00539 & $\pm$ 0.00845 & $\pm$ 0.01068\\
        $K=32$  & 13.04421  & 13.07223  & 13.123387 \\
        & $\pm$ 0.00508 & $\pm$ 0.00512 & $\pm$ 0.01180\\
        $K=64$  & 13.03640  & 13.08978  &13.130018 \\
        & $\pm$ 0.00266 & $\pm$ 0.01689 & $\pm$ 0.00835 \\
        $K=128$ & 13.035136  & 13.07355  & 13.155595  \\
        & $\pm$ 0.00369 & $\pm$ 0.00572 & $\pm$ 0.02530 \\
        \\
        No deconvolution  & 13.10600 & 13.1185 & 13.1601\\\\
        Ground truth &  13.01441 & 13.0500 & 13.0987\\
    \end{tabular}
    \label{tab:6dlossratio}
\end{table}

\subsubsection{Varying the number of bins in the normalizing flow architecture}\label{sec::varying_B}

As discussed in Section~\ref{architecture}, there are several hyperparameters in the rational quadratic spline transform that one must select to produce a faithful representation of the underlying distribution. We have found the most significant of these is the bin number $B$, which controls the complexity and flexibility of each layer of the flow model. Higher $B$ allows the model to capture more subtle features and generate flow models with finer resolution as more bins are used in the transform function. For the output layer, $B$ controls the number of knots in the non-equally-spaced splines and so can be considered very approximately as a measure of the average resolution across each dimension. An overly flexible model can produce artefacts that do not exist in the `ground truth' distribution. 

We produce such artefacts when using a large bin number, $B = 32$, and a very small data set, $\eta = 0.01$. Given the autoregressive and spline nature of the flow model, the artefacts manifest as gridded features as shown in Fig.~\ref{fig:0.01-spiral}. A similar plot for the velocity distribution is shown in Fig.~\ref{fig:0.01-velocity}. These artefacts are irrelative to the sample number $K$. The gridded artefacts have only shown up visually when oversampling a model with $\sim500,000$ samples trained with a smaller dataset of only $\sim7000$ stars. If one visualizes only $7000$ samples from the model, the artefacts are not perceptible. This behaviour is a natural consequence of an overly flexible model unconstrained by any priors or regularization. As discussed in Section~\ref{method}, the deconvolution is well-determined, but there is a range of models consistent with the Poisson sampling of the underlying distribution. 

\begin{figure}
    \centering
    \includegraphics[width=0.481\textwidth]{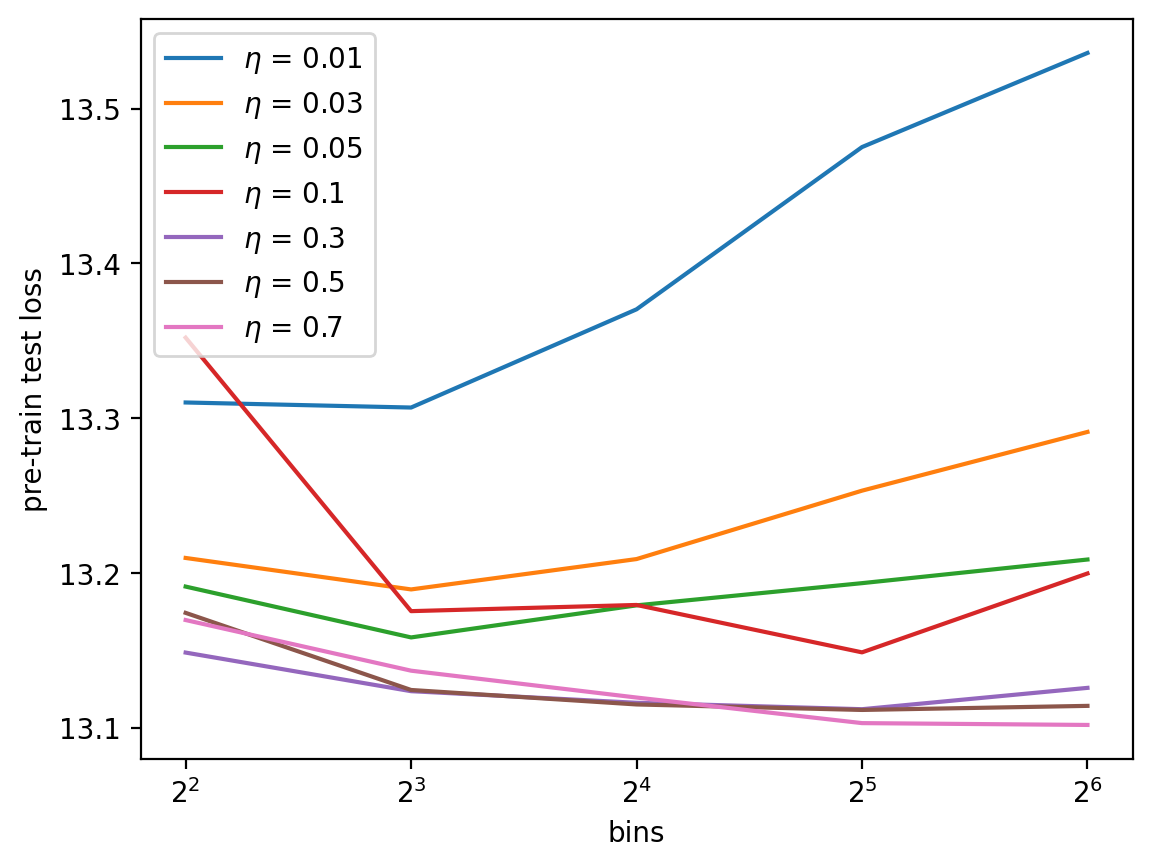}
    \caption{Test loss against model bin number $B$ after 100 epochs of the pre-train stage. $\eta$ is the relative train data size. The test loss of models with $\eta \lesssim 0.1$ show a clear `V' shape that indicative of over-smoothing at small $B$ and an overly flexible model at large $B$. For $\eta \gtrsim 0.1$, the minimum/optimum $B$ shifts higher, showing that the quality of the data merits a higher resolution model. }
    \label{fig:pre-train}
\end{figure}

To reduce the impact of these features, one must introduce regularization or priors. One effective way of doing this is reducing the bin number $B$ for the spline layer, which smooths the flow model by decreasing resolution and hence removes these unphysical artefacts. Tuning other hyper-parameters, such as the number of autoregressive layers $L$, hidden neural network size $h$, or an L2 regularisation factor $\omega$ do not prove as effective in improving the fitting. However, simply reducing $B$ is an imprecise tool as there is a risk of over-smoothing and not denoising the distribution at all. Unfortunately, due to the very flexible nature of the model, it is hard to compute an effective resolution given $B$, so it seems the only robust way to use the models in this generalised sense is to perform detailed cross-validation. Fig.~\ref{fig:pre-train} shows the test loss versus the bin number $B$ of the models after 100 epochs of the pre-train stage using different data sizes, $\eta$. With a small data set with $\eta < 0.1$, the test loss shows a clear `V' shape: for low $B$ the models are over-smoothed whilst for large $B$ the models are overly flexible leading to an optimum $B\approx8$. When $\eta\gtrsim0.1$, the `V' shape is less obvious but the plotsshows that the optimum $B$ shifts higher to $B\approx32$ (as used in the sections above). 

\section{Discussion}\label{sec:discussion}
We have provided a proof of principle for our proposed methodology and demonstrated the method's power in recovering subtle data features. We now critically compare our approach with other approaches, most notably extreme deconvolution,  and then discuss how our models can be utilised in practical applications.

\subsection{Other normalizing flow architectures}\label{discussion-architectures}
As mentioned in Section~\ref{architecture}, we chose the autoregressive rational quadratic spline flow in our work. However, our approach could easily be switched to using any available flow architecture. We have also tested our model with other flow architectures such as Masked Autoregressive Flow \citep{papamakarios2018maf}, Gaussianization Flows \citep{meng2020gaussianizationflows}, Bernstein-Polynomial Normalizing Flows \citep{sick2020bernstein-polynomial} and continuous flow models such as FFJORD \citep{grathwohl2018ffjordfreeformcontinuousdynamics}. Among those flow architectures, rational quadratic spline flow converges fast enough to balance the computational cost and robustness requirement. One disadvantage of this architecture is the gridded artefacts as discussed in Section~\ref{sec::varying_B}, which could be avoided by carefully tuning the flow model in the pre-train stage.

\subsection{Comparison with extreme deconvolution}\label{sec:xdgmm}

\begin{table}
    \caption{Test loss per star for XDGMM models of the $q=5$ $\eta=0.7$ dataset and GMM models of the ground truth dataset with varying numbers of Gaussian components $N$.}
    \centering
    \begin{tabular}{|c|c|c|}
        &\multicolumn{2}{c}{Test loss}\\
        Setup & $q=5$ XDGMM & Ground truth GMM\\
        \hline
        $N=32$  & $13.17224 \pm 0.00407$ & 13.27508\\
        $N=64$  & $13.13887 \pm 0.00175$ & 13.17699\\
        $N=128$  & $13.12934 \pm 0.00369$ & 13.13118\\
        $N=256$  & $13.12398 \pm 0.00355$ & 13.13045\\
        $N=512$  & $13.11939 \pm 0.00100$ & 13.10757\\
        $N=1024$  & $13.11641 \pm 0.00163$  & 13.09030\\
        $N=2048$  & $13.11519 \pm 0.00171$  & 13.12894\\
        \\
    \end{tabular}
    \label{tab:loss_gmm}
\end{table}

\begin{figure}
    \centering
    \includegraphics[width=0.481\textwidth]{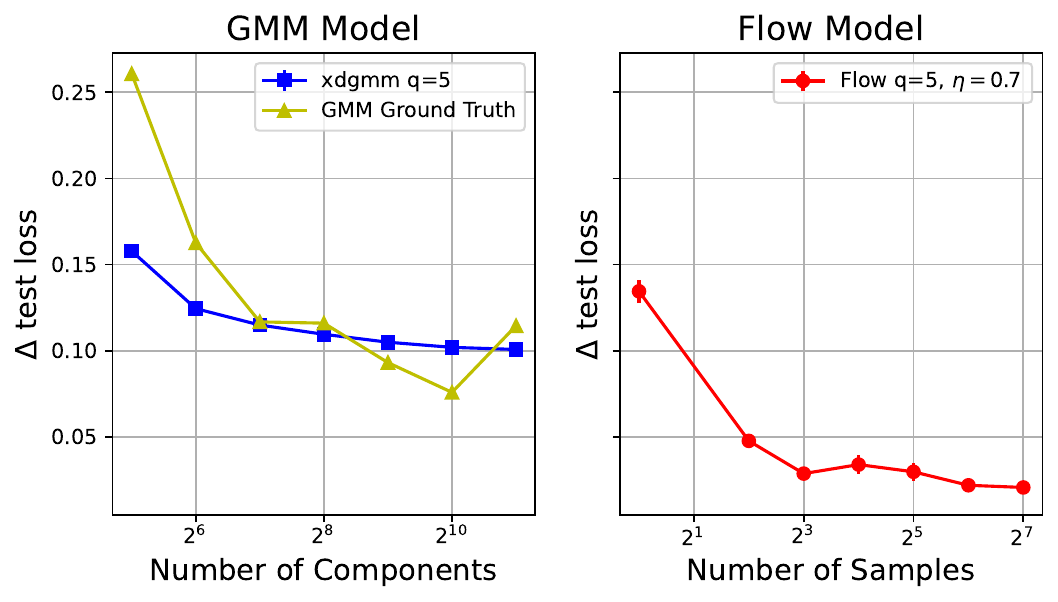}
    \caption{Comparison of the test loss per star for the (XD)GMM models (left, plotted against number of Gaussian components) and normalizing flow models (right, plotted against number of denoising samples, $K$). In the left plot, the yellow triangles show results for a GMM fit to the noise-free sample whilst the blue squares show the results for a full XDGMM fit using the $q=5$ training set. In both panels, the test loss per star is reported relative to the `best-case scenario' using a flow model fitted to the noise-free data with $\eta = 0.7$.}
    \label{fig:test_loss_gmm}
\end{figure}

\begin{figure*}
    \centering
    \includegraphics[width=\textwidth]{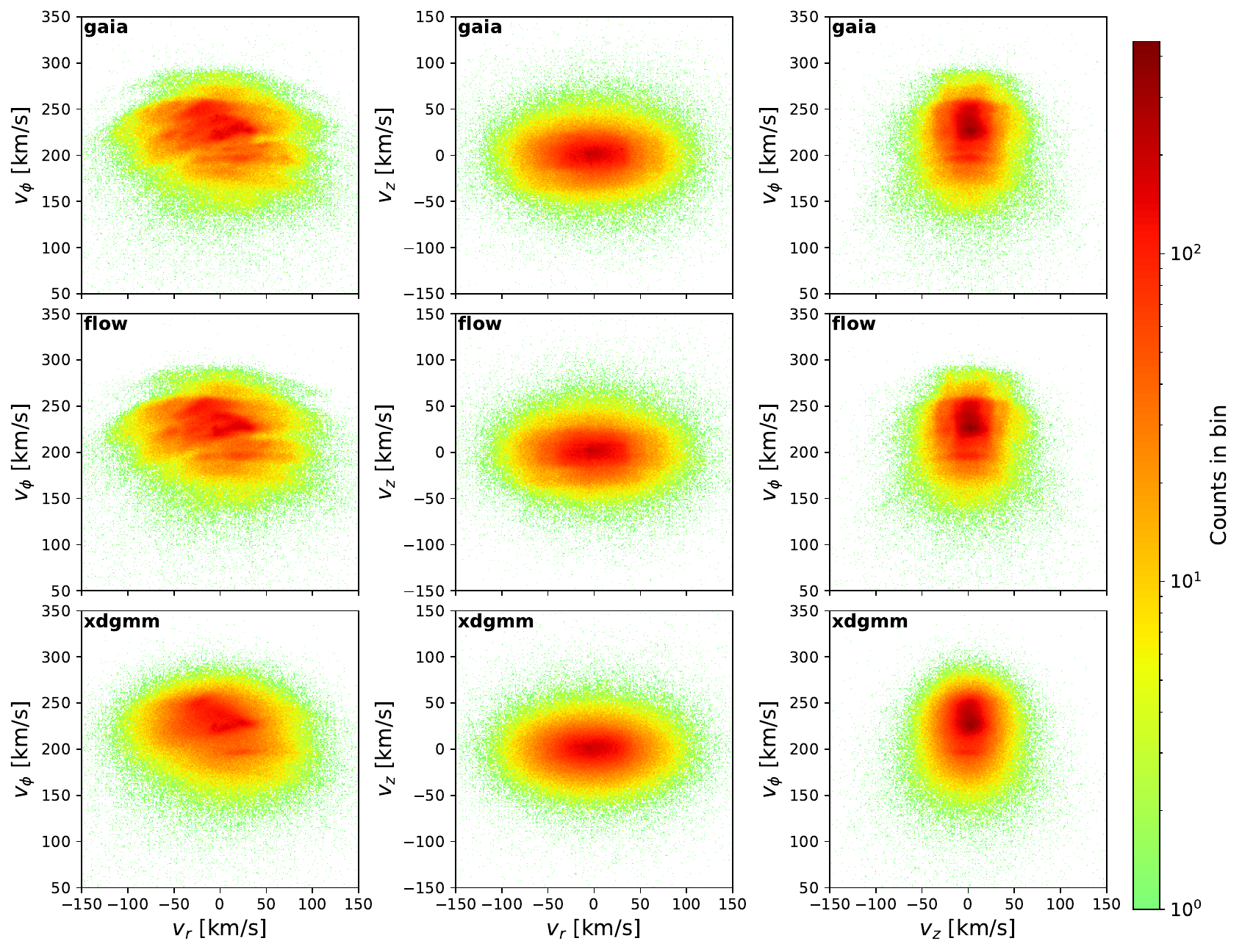}
    \caption{Comparison of the model XDGMM and normalizing flow velocity distributions within the Galactic height range $|z|<2 \mathrm{kpc}$. 
    The top row is the `ground-truth' \emph{Gaia} test data set, the second row is the samples from the flow model using $q=5$ and $\eta=0.7$ and the last row is the samples from the fitted XDGMM with $N=1024$ (again using $q=5$ and $\eta=0.7$).     
    \begin{inparaenum}
    \protect\item in the left column, although the XDGMM model recovers part of the Hercules stream arches, it fails to recover the boundaries between the Hercules stream arches at $v_{\phi} = 180$ km/s and other moving groups at $v_{\phi} = 220$ km/s. The energy boundary at $v_{\phi} = 280$ km/s becomes fuzzy.
    \protect\item In the middle column, the `shell' at  $v_z$ = -30 km/s is not enhanced from the background with XDGMM model.
    \protect\item In the right column, the XDGMM model shows a narrow arch feature sitting in a smoother background when compared with Gaia data and flow model.
    \end{inparaenum}
    }
    \label{fig:6d_v_gmm}
\end{figure*}

The main alternative to our method in the literature is extreme deconvolution \citep[XDGMM][]{Bovy_2011}. XDGMM models the distribution as a Gaussian mixture model but accounts for Gaussian uncertainties during training. We provide benchmark results with standard GMM model fitting through the \textsc{sklearn} package and an upgraded version of XDGMM, Scalable Extreme Deconvolution \citep{ritchie2019scalable}, supported by the GPU acceleration algorithm. \citet{2024arXiv241203029K} generalise the XDGMM method by using a neural network over a conditional $c$ to obtain the GMM model's mixing coefficient, mean and covariance matrix. However, just like other GMM methods, it requires the target distribution to be an approximate Gaussian mixture. Standard GMM models are trained with the ground truth set-up in Section~\ref{sec:corruption} to compare the expressive power between the flow method and the Gaussian method mixture. Similarly, XDGMM models are trained with the corrupted mock dataset with $q=5$ and $\eta = 0.7$ to compare the denoising ability.

The XDGMM method requires the covariance matrices $\Sigma _\mathrm{gc}$ for each star in the z-scored Galactocentric coordinate. We obtain these from the covariance of $100$ samples generated from the observable uncertainty distribution and propagated to Galactocentric coordinates. We set the off-diagonal terms in the resulting covariance matrices to zero and add a small value $\epsilon = 2\times10^{-5}$ to the diagonal terms to improve the numerical stability. Note that, unlike our importance sampling approach which correctly captures the full uncertainty distribution in the modelled (Galactocentric) space, the XDGMM approach requires us to only use Gaussian approximations to the error distribution. Most of the hyper-parameters for XDGMM are set to their defaults. Further to this, we set the number of $\texttt{epochs} = 20$, the regularisation constant $w = 0.001$, $\texttt{batch size} = 512$ and initial learning rate $\texttt{lr}= 0.01$. After $10$ epochs, the learning rate reduces by a factor of $10$.

The results of the GMM model of the ground truth data and the XDGMM model of the mock dataset using different numbers of components $N$ are shown in Table~\ref{tab:loss_gmm} and Fig.~\ref{fig:test_loss_gmm}. As shown in Fig.~\ref{fig:test_loss_gmm}, the test loss decreases as $N$ increases for both GMM models and converges when $N \approx 1024$. Both GMM models struggle with this complicated 6D dataset. The flow model outperforms them when the number of denoising samples $K = 4$. The velocity distribution of the $N=1024$ XDGMM model is shown in Fig.~\ref{fig:6d_v_gmm}. It is blurrier compared to the flow model's distribution, and the Hercules stream is transformed into one narrow arch of overdensity on top of a broad background. These comparisons show the flow model's advantages of more expressive power and flexibility, and the potential for the flow model to recover small-scale features that (XD)GMM methods fail to.

The number of parameters used in the flow model scales with the dimension $D$ and bin number $B$ as $O(DB)$. For our model, the number of parameters is $545628$. In comparison, the (XD)GMM model with $2048$ components has $57344$ parameters. Even though the flow model has an order of magnitude more parameters, the size of the model is still acceptable. Reducing the bin number $B$ can significantly decrease the model's size for simple tasks that do not require high-resolution features but as already described, the relationship between $B$ and resolving power is non-trivial.

One further consideration when comparing the algorithms is the training time. The typical training time of the normalizing flow models varies as the number of denoising samples, $K$, increases. We use a single Nvidia Tesla T4 GPU for our training (via Google Colab). As expected, the training time is approximately linear with $K$: for $K = 32$, it takes 4 hours to train the flow, whilst for $K = 128$, it takes 16 hours to train the flow. It takes approximately 1 hour to train an XDGMM model with 2048 Gaussian components using the GPU accelerated algorithm from \citep{ritchie2019scalable}. The normalising flow method takes longer to train, but we advocate that it is the only reasonable choice for resolving high-resolution features.

\subsection{Utility of models}
Our models offer significant utility, particularly in \emph{Gaia} data analysis pipelines. The flow model's denoising ability can be used to provide a noise-free upgrade to the existing application of normalizing flow e.g. in equilibrium model fits \citep{green2020deep,Green_2023,kalda2023recovering}. It can also extract solar neighbourhood dynamical features possibly related to bar or spiral resonances or obtain a distribution function for halo stars, a regime where uncertainties can become significant. As described, our approach only allows for the estimation of the underlying distribution in the absence of any observational noise. Typically in Galactic applications, this distribution is not purely the distribution of the stars in the Milky Way but is affected by observational limitations (on-sky, depth, extinction etc.). The combination of these effects is often called the selection function. We imagine these effects can mostly be modelled \emph{after} the denoising process, although we note that it is possible data selection is performed on noisy measurements and it may not be possible to separate these steps. In this case, our denoising procedure must be incorporated as part of a fuller modelling framework.

A denoised model is also a more robust route to projecting the data into more complex spaces, e.g. action-angle space. Uncertainties transform in significantly non-linear ways when projecting to action-angle space, e.g. a star on a near-circular orbit can easily move from pericentre to apocentre of their orbit with a small shift in velocity, so simply projecting the data can give rise to artificial features. By first denoising flexibly, we can project the `true' distribution to these more complex spaces and give a more faithful representation. Similarly, we believe our models act as useful surrogate models that can be fed into more physically motivated inference models. An analysis would then consist of two steps: flexible denoising to generate a surrogate model followed by the comparison of e.g. a galaxy model to the denoised surrogate model. Finally, it is likely our models can be utilised as an information compression tool for easier and more lightweight sharing and comparison of the \emph{Gaia} data. A normalizing flow model is a useful summary tool that can be plotted at any required resolution and projection, and does not require the manipulation of the full \emph{Gaia} data.

\section{Conclusions}\label{sec:conclusions}
The interpretation of Milky Way stellar survey data from e.g. \emph{Gaia} often requires density deconvolution in the presence of heteroscedastic noise to recover features indicative of the structure and history of the Galaxy. We have outlined a practical approach to using normalizing flows for this problem, and demonstrated the success in reproducing fine features in the solar neighbourhood such as the multiple branches of the Hercules stream and the \emph{Gaia} phase-space spiral. We have discussed how different choices of hyperparameters and the training set size change the results and give appropriate choices for our specific use case. Our results are a proof of principle that the proposed methodology is useful for \emph{Gaia} data analysis, but we envisage immediate utility in applying the method to noisy \emph{Gaia} data samples well beyond the solar neighbourhood. Furthermore, the proposed approach will be useful for
\begin{inparaenum}
\item transformation to richer spaces, e.g. action space, where noise properties distort the data in non-linear ways,
\item surrogate models that have flexibly captured properties of noise but can be directly compared with more physically motivated models,
\item convenient information compression for lightweight sharing of the \emph{Gaia} results.
\end{inparaenum}

\section*{Acknowledgements}
We use the \textsc{Zuko} flow library \citep{2023zndo...7625672R}  to construct the normalizing flow models used in this paper. We also made use of \textsc{PyTorch} \citep{2019arXiv191201703P}, \textsc{Astropy} \cite{2022ApJ...935..167A}, \textsc{Matplotlib} \citep{2007CSE.....9...90H} and \textsc{Numpy} \citep{2020Natur.585..357H}.

We thank Jason Hunt for useful feedback on the manuscript.

JLS acknowledges support from the Royal Society (URF\textbackslash R1\textbackslash191555). This work has made use of data from the European Space Agency (ESA) mission {\it Gaia} (\url{https://www.cosmos.esa.int/gaia}), processed by the {\it Gaia} Data Processing and Analysis Consortium (DPAC, \url{https://www.cosmos.esa.int/web/gaia/dpac/consortium}). Funding for the DPAC has been provided by national institutions, in particular the institutions participating in the {\it Gaia} Multilateral Agreement.

%%%%%%%%%%%%%%%%%%%%%%%%%%%%%%%%%%%%%%%%%%%%%%%%%%
\section*{Data Availability}
All data used in this work are in the public domain. The code used is in this paper is available: \url{https://github.com/Ziyang-Yan/Denoising-Milky-Way-with-NF.git}



\bibliographystyle{mnras}
\bibliography{Bibliography} % if your bibtex file is called example.bib

%%%%%%%%%%%%%%%%%%%%%%%%%%%%%%%%%%%%%%%%%%%%%%%%%%

%%%%%%%%%%%%%%%%% APPENDICES %%%%%%%%%%%%%%%%%%%%%

\appendix
\section{Additional figures}
In this appendix, we provide some additional supporting figures.
\subsection{Residual plot for q=5 model}\label{residual}
\begin{figure*}
    \centering
    \includegraphics[width=\textwidth]{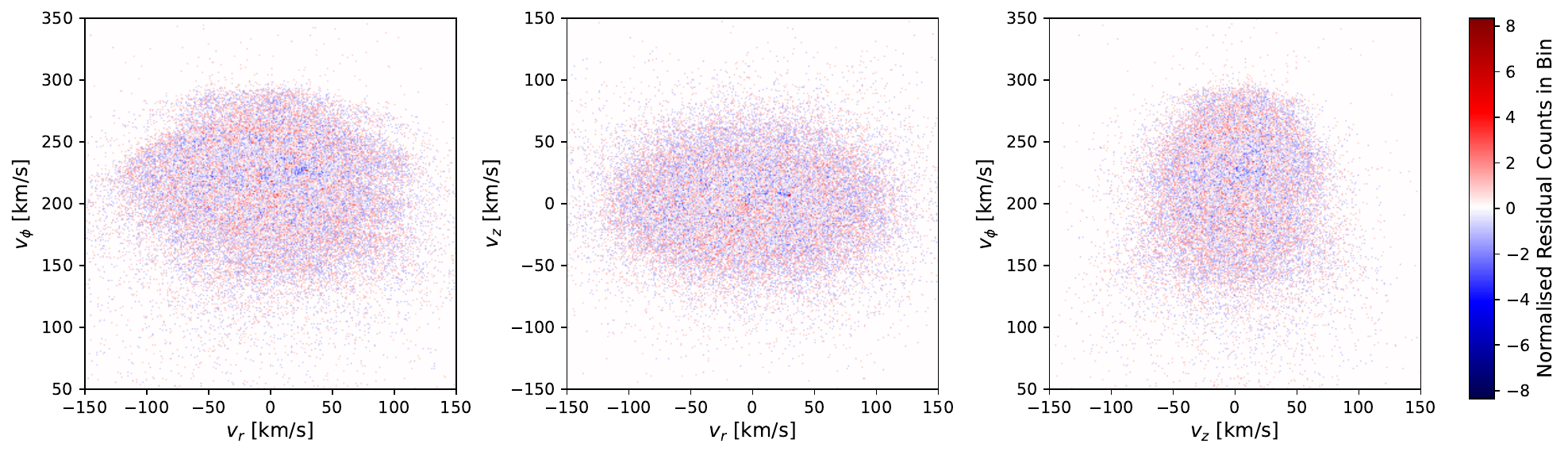}
    \caption{The residual plot of Gaia data and the $q=5$ flow model samples, which is calculated by the z-scored difference bin counts of \ref{fig:6d_v_0.05_amp5}. More specifically, $(N_{Gaia} - N_{flow}) / \sqrt{N_{Gaia}} $ or $(- N_{flow}) / \sqrt{N_{flow}}$ if $N_{Gaia} = 0$.
    }
    \label{fig:residual}
\end{figure*}

Fig.~\ref{fig:residual} shows the normalised residual plot between the Gaia data and the samples of the $q=5$ flow model.

\subsection{Result for q=10 model} \label{amp10}
\begin{figure*}
    \centering
    \includegraphics[width=\textwidth]{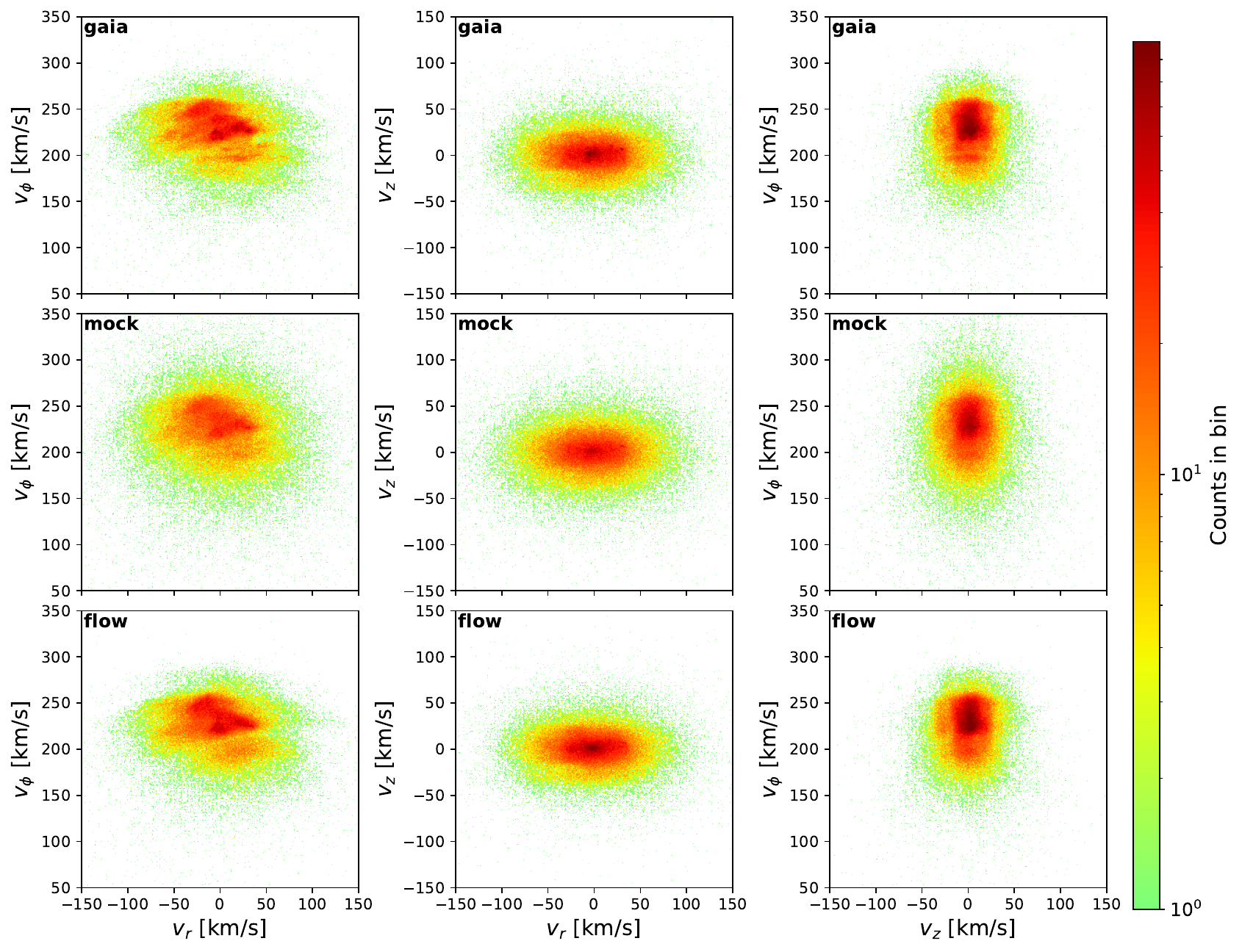}
    \caption{As per Fig.~\ref{fig:6d_v_0.05_amp5} but for the $q=10$ training set. In this case, we note:
    \begin{inparaenum}
    \protect\item In the left column, similar to the model for the $q=5$ training set, the flow model enhances the moving group and the energy boundary at $v_{\phi} = 280\,\mathrm{km/s}$, but in this case, fails to recover the boundary between the Hercules Stream arches and only separates that region out as a whole;
    \protect\item In the middle column, the flow model recovers the `shell' at  $v_z$ = -30 km/s as for the $q=5$ test set.
    \protect\item In the right column, the flow model constrains the distribution in the central region but fails to recover the arch feature at $v_{\phi}$ = 200 km/s.
    \end{inparaenum}
    }
    \label{fig:6d_v_0.05_amp10}
\end{figure*}

\begin{figure*}
    \centering
     \includegraphics[width=0.9\textwidth]{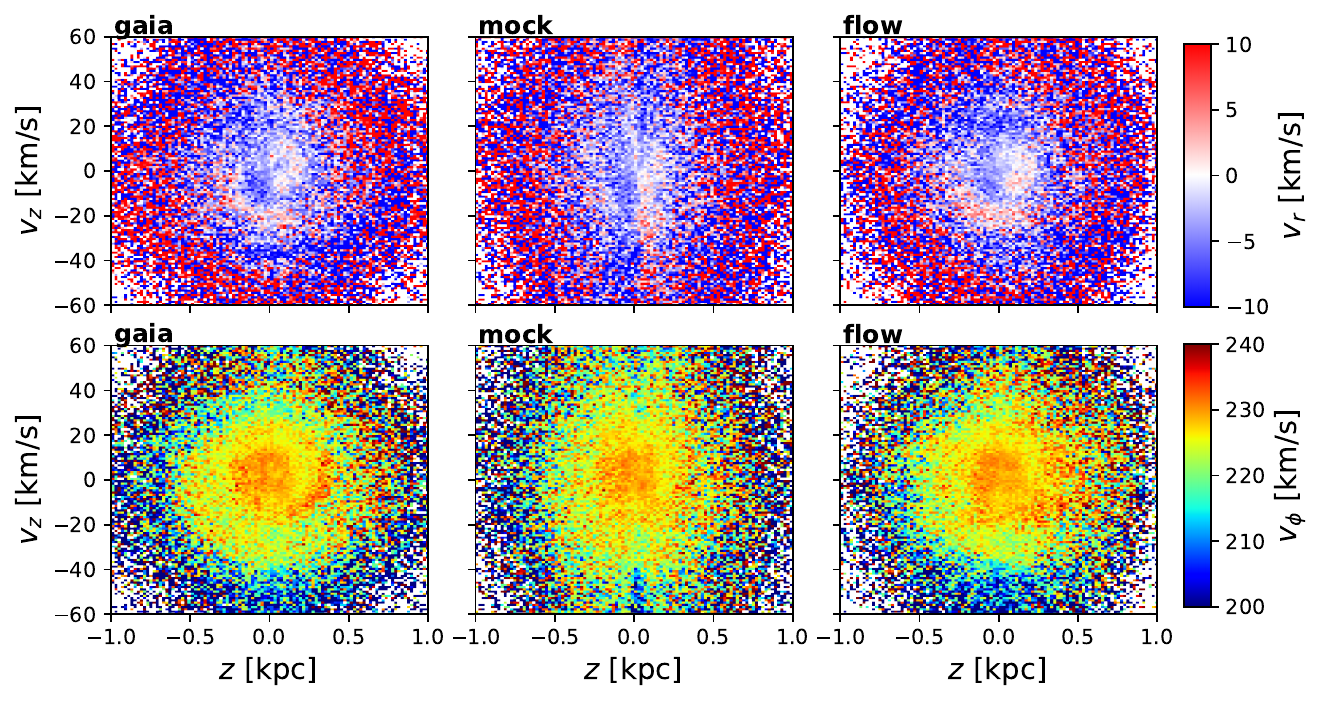}
    \caption{Phase spiral recovery using the 6d model on the $q=10$ training set (see Fig.~\ref{fig:6d_spiral_amp5} for a description of the different panels). The flow model corrects the distorted feature at the plane ($z=0$) yet fails to recover the clear spiral pattern.}
    \label{fig:6d_spiral_amp10}
\end{figure*}

In the main body of the paper, we inspected qualitatively the performance of our procedure using mock data scattered by $q=5$ times the reported \emph{Gaia} uncertainties. Here we present equivalent results using a more extreme case of $q=10$. Fig.~\ref{fig:6d_v_0.05_amp10} shows the velocity distribution of the models fitted to the $q=10$ training set. The model only enhances the boundary between the Hercules stream and the bulk of disc stars; it does not recover the detailed features of the Hercules stream. This is a generic issue: although the deconvolution is well-posed, the Poisson error from finite sampling is such that there are many deconvolved distributions that are consistent with the target data. As shown in Fig.~\ref{fig:6d_spiral_amp10}, the model for the $q=10$ training set fails to recover all the spiral features, but still successfully removes the noise at the Galactic plane. 

\subsection{Full 6D corner plot}
\begin{figure*}
    \centering
    \includegraphics[width=\textwidth]{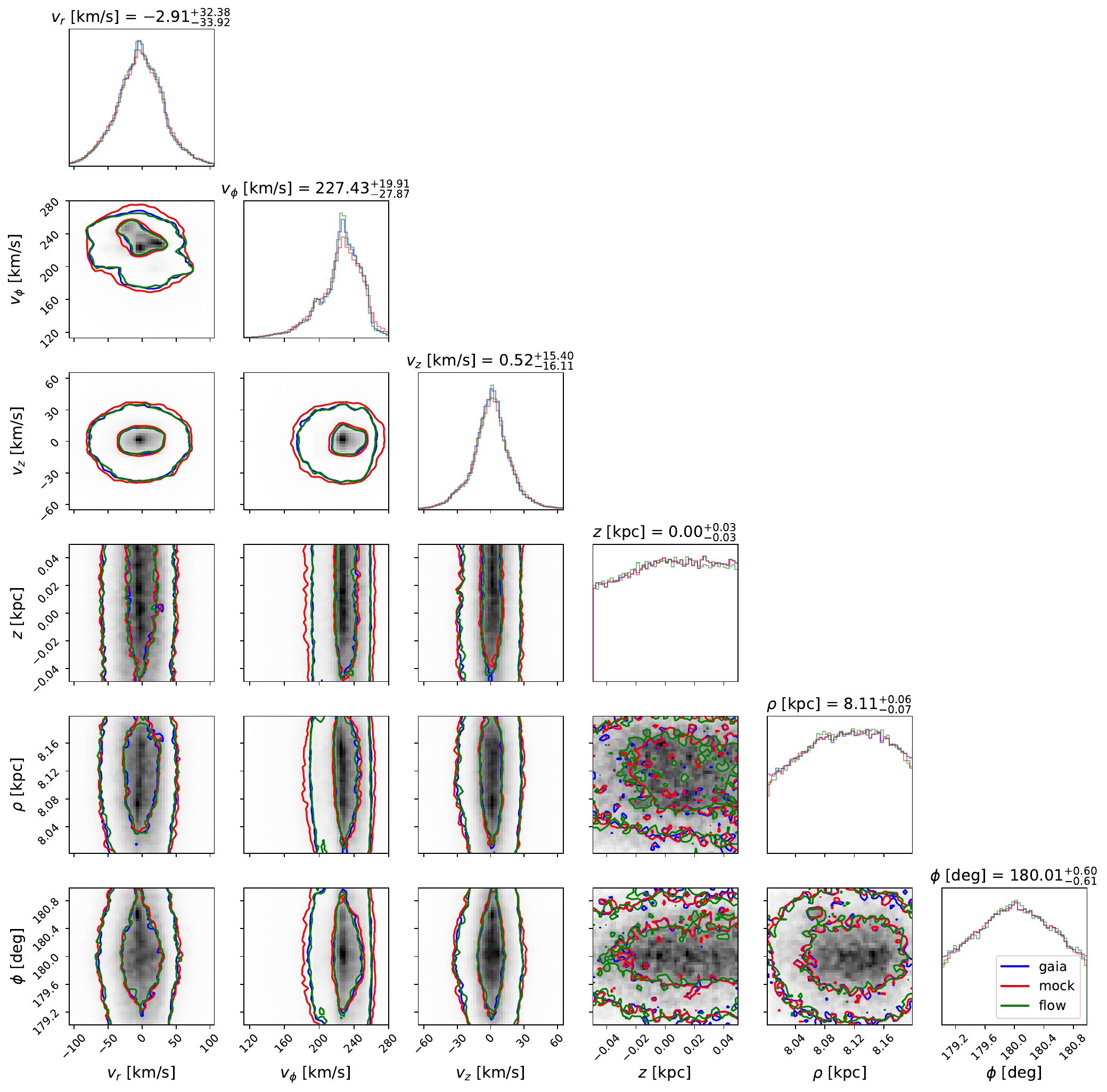}
    \caption{Corner plot for the 6d model fitted to the $\eta = 0.7,q=5$ dataset (shaded density). Overplotted are contours for the original Gaia data (blue), the corrupted mock data (red) and flow samples (green). The statistics at the top of each diagonal panel are obtained from the Gaia data. The plot shows 99\% of the data in each dimension. Two contours represent one sigma (containing 39.3\% of the sample) and two sigma (containing 86.4\% of the sample).}
    \label{fig:corner_amp5}
\end{figure*}

\begin{figure*}
    \centering
    \includegraphics[width=\textwidth]{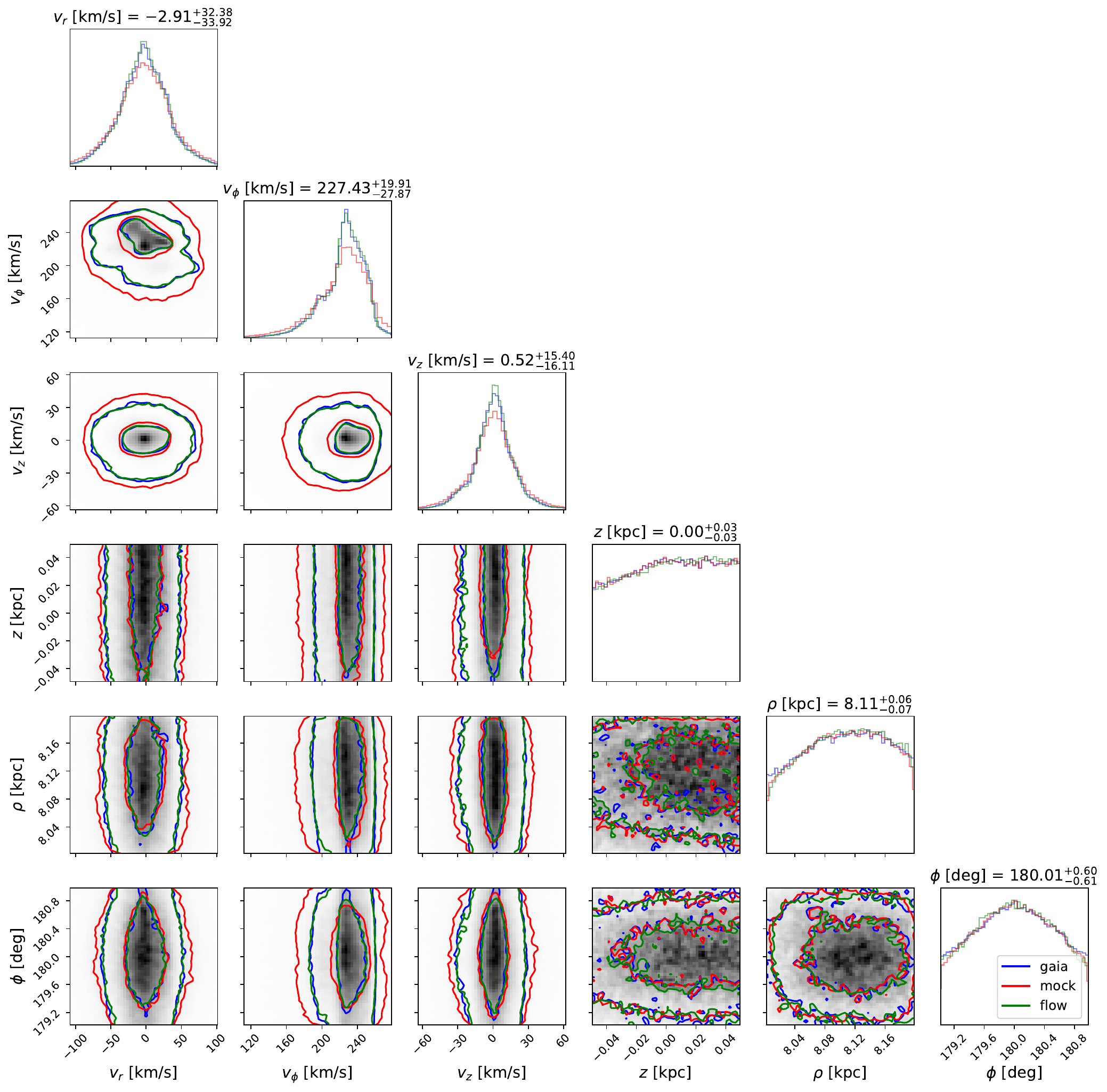}
    \caption{As Fig.~\protect\ref{fig:corner_amp5} but for the model applied to the $\eta = 0.7,q=10$ training set.}
    \label{fig:corner_amp10}
\end{figure*}

We show full 6D corner plots for the noise-free data, mock data and denoised normalizing flow models in Fig.~\ref{fig:corner_amp5} for the $q=5$ mock dataset and Fig.~\ref{fig:corner_amp10} for the $q=10$ mock dataset.

%%%%%%%%%%%%%%%%%%%%%%%%%%%%%%%%%%%%%%%%%%%%%%%%%%

% Don't change these lines
\bsp	% typesetting comment
\label{lastpage}
\end{document}